# Comparative Autoignition Trends in the Butanol Isomers at Elevated Pressure

*Bryan W. Weber\* and Chih-Jen Sung*

Department of Mechanical Engineering, University of Connecticut, Storrs, CT 06269, USA

Corresponding Author:

Bryan W. Weber

Phone: 860-486-2492

Email: bryan.weber@uconn.edu

# Comparative Autoignition Trends in the Butanol Isomers at Elevated Pressure

*Bryan W. Weber and Chih-Jen Sung*

Department of Mechanical Engineering, University of Connecticut, Storrs, CT 06269, USA

**Abstract**

Autoignition experiments of stoichiometric mixtures of *s*-, *t*-, and *i*-butanol in air have been performed using a heated rapid compression machine (RCM). At compressed pressures of 15 and 30 bar and for compressed temperatures in the range of 715−910 K, no evidence of a negative temperature coefficient region in terms of ignition delay response is found. The present experimental results are also compared with previously reported RCM data of *n*-butanol in air. The order of reactivity of the butanols is *n*-butanol>*s*-butanol≈*i*-butanol>*t*-butanol at the lower pressure, but changes to *n*-butanol>*t*-butanol>*s*-butanol>*i*-butanol at higher pressure. In addition, *t*-butanol shows pre-ignition heat release behavior, which is especially evident at higher pressures. To help identify the controlling chemistry leading to this pre-ignition heat release, off-stoichiometric experiments are further performed at 30 bar compressed pressure, for *t*-butanol at $\phi = 0.5$ and $\phi = 2.0$ in air. For these experiments, higher fuel loading (i.e. $\phi = 2.0$) causes greater pre-ignition heat release (as indicated by greater pressure rise) than the stoichiometric or $\phi = 0.5$ cases. Comparison of the experimental ignition delays with the simulated results using two literature kinetic mechanisms shows generally good agreement, and one mechanism is further used to explore and compare the fuel decomposition pathways of the butanol isomers. Using this mechanism, the importance of peroxy chemistry in the autoignition of the butanol isomers is highlighted and discussed.

# 1. Introduction

Petroleum has long been the source of fuels used in many applications, from stationary power generation to transportation systems. With the world's oil supply in geo-political uncertainty and continuing climate issues, there is a renewed effort to develop a replacement for petroleum, particularly for use in transportation applications. Ideally, this replacement should offer many of the benefits of petroleum based fuels – easy distribution, high energy density, and low cost – while eliminating the disadvantages – impact on climate change and uncertain supply sources.

The most commonly proposed fuel sources recently have been biomass sources. The fuels made from these renewable biological sources offer many of the desired properties of a new fuel – easy distribution, carbon-neutral, local production – but current generation bio-derived fuels (or biofuels) still suffer from some of the disadvantages of petroleum. A typical example of a biofuel is ethanol, which has been used as an additive in gasoline for many years. Another example is bio-diesel, which is typically made from plant oils rather than oil removed from the ground. Although ethanol is ubiquitous at gasoline pumps, it suffers from several disadvantages that suggest it needs to be replaced.[1] In particular, ethanol has a much lower energy density than gasoline, reducing volumetric fuel economy, and ethanol is typically produced from crops that would otherwise be used as food sources.[2]

$n$-Butanol has recently been identified as one of a suite of so called "second generation" biofuels intended to supplement or replace the "first generation" biofuels currently in use, such as ethanol.[3,4] The second generation biofuels will help alleviate some of the problems identified with the first generation biofuels, including concerns about feedstocks. In addition to the normal ($n$) isomer, there are three other isomers of butanol – $s$-, $i$-, and $t$-butanol. Biological production pathways have been identified for *n*-, *s*-, and *i*-butanol,[4,5] but *t*-butanol is a petroleum derived product. Nevertheless, *t*-butanol is currently used as an octane enhancer in gasoline.

In the last five years, research into the combustion characteristics of the isomers of butanol has exploded. It is therefore not our intention to provide a comprehensive overview of the literature; rather, we will only provide exemplary references except for the articles of particular interest to this study. In addition to applied engine research,[6–8] fundamental combustion measurements have been made using many different systems. These include laminar flame speeds,[9] flame stability measurements,[10] jet-stirred reactor chemistry,[11] low-pressure flame structure,[12] atmospheric pressure flame structure,[13] pyrolysis,[14] flow reactors,[15] and ignition delays, both in rapid compression machines (RCMs) and shock tubes.[16,17] There have also been numerous modeling studies of the isomers of butanol, including recent ones by Sarathy et al.,[18] Hansen et al.,[19] and Merchant et al.[20]

Due to its relevance in predicting the performance of a fuel in existing and advanced engines, ignition delay is a very common measure of the global performance of a kinetic mechanism. Ignition delays for homogeneous systems are typically measured in shock tubes or RCMs, where the effects of fluid motion and turbulence are generally minimized. Several studies of ignition delay of the butanol isomers have been conducted in both apparatuses, including work in shock tubes by Black et al.,[21] Heufer et al.,[22] Moss et al.,[23] Noorani et al.,[24] Stranic et al.,[17] Vasu et al.,[25] and Vranckx et al.,[26] and work in RCMs by Karwat et al.[27] and Weber et al.[16] These studies have covered a wide range of temperature-pressure regimes for *n*-butanol, from 1–90 bar and 675–1800 K. However, only the shock tube studies from Moss et al.[23] and Stranic et al.[17] have studied all four isomers of butanol, over the more limited range of 1–48 atm and 1022–1825 K.

Despite the intense work in the recent past to generate fundamental combustion data for the isomers of butanol, there is still a limited amount of published work for low-temperature and high-pressure conditions. The goal of this study is to provide additional autoignition data for the three isomers of butanol besides *n*-butanol at elevated pressure and low-to-intermediate temperature; this work is intended as a companion to the recent work on *n*-butanol published by our group.[16] Using the same rapid compression machine as the previous study, we have characterized the ignition delays of *s*-, *i*-, and *t*-butanol for pressures from 15 to 30 bar and temperatures in the range of 715–910 K.

## 2. Experimental and Computational Specifications

### 2.1 Rapid Compression Machine

Autoignition delay experiments have been performed in a heated rapid compression machine (RCM). The RCM compresses a fixed mass of premixed fuel and oxidizer to a given pressure and temperature over about 30–40 ms. The volumetric compression ratio, initial temperature, and initial pressure are adjusted to vary the top dead center (TDC) temperature at a constant TDC pressure. The piston in the reaction chamber is machined with specifically designed crevices to ensure that the roll-up vortex effect is suppressed and homogeneous conditions in the reaction chamber are promoted.

The reaction chamber is fitted with a Kistler 6125B dynamic pressure transducer to measure the pressure during and after compression, including any post-compression events such as ignition. In addition, an Omega Engineering PX303 static pressure transducer and a K-type thermocouple are installed to measure the pre-compression pressure and temperature, respectively. Further details of the experimental setup can be found in the work by Mittal and Sung.[28]

## 2.2 Mixture Preparation

The reactants used in this study, along with their purities, are shown in Table 1. To determine the relative proportions of each reactant in the mixture, the absolute mass of fuel, the equivalence ratio ($\phi$), and the oxidizer ratio ($X_{O_2}:X_{inert}$, where $X$ indicates mole fraction) are specified. *s*- and *i*-Butanol are liquid at room temperature and have relatively low vapor pressure; therefore, each is measured gravimetrically in a syringe to within 0.01 g of the specified value. *t*-Butanol is solid at room temperature (melting point: 25 °C), and is melted before being handled in the same procedure as the other fuels. The 17 L mixing tank is vacuumed to an ultimate pressure less than 5 Torr prior to the injection of the liquid fuel through a septum. Proportions of $O_2$ and $N_2$ are added manometrically at room temperature. The preheat temperature of the RCM is set above the saturation point for each fuel to ensure complete vaporization. A magnetic stirrer mixes the reactants. The temperature inside the mixing tank is allowed to equilibrate for approximately 1.5 hours.

This approach to mixture preparation has been validated in several previous studies by withdrawing gas samples from the mixing tank and analyzing the contents by GC/MS, GC-FID, and GC-TCD.[16,29,30] These studies have verified the concentration of *n*-butanol, water, and *n*-decane, respectively. In addition, both the work by Kumar et al.[30] on *n*-decane and the study of Weber et al.[16] on *n*-butanol confirmed that there was no fuel decomposition over the course of a typical set of experiments. Furthermore, within this study, each new mixture preparation is checked against previously tested conditions to ensure reproducibility.

## 2.3 Experimental Conditions and Repeatability

Table 1 shows the experimental conditions considered in this study. The compressed pressure conditions have been chosen to match the previous *n*-butanol study,[16] but also to provide data in regions not covered extensively in previous work. In addition, the fuel loading conditions have been chosen to complement previous work; the studies by Stranic et al.[17] and Moss et al.[23] used relatively dilute mixtures, so we have included higher fuel loading conditions. Furthermore, the compressed temperature conditions we have studied ($T_C = 715 - 910$ K) have not been examined in any other study, to our knowledge.

Each compressed pressure and temperature condition is repeated at least six times to ensure repeatability. The mean and standard deviation of the ignition delay for all runs at each condition are calculated. As an indication of repeatability, the standard deviation is less than 10% of the mean in every case. Representative experimental pressure traces for simulations and presentation are then chosen as the closest to the mean. A complete list of the experimental conditions, including mixture composition, initial temperature and pressure, compressed temperature and pressure, and ignition delay results, is available in the Supporting Information.

## 2.4 Simulations and Determination of Compressed Temperature

Two types of simulations are performed using CHEMKIN-Pro.[31] The first type of simulation is a constant volume, adiabatic simulation, whose initial conditions are set to the TDC pressure and temperature conditions determined in the RCM experiments. The second type includes both the compression stroke and post-compression event by setting the simulated reactor volume profile as a function of time. These are respectively denoted as CONV and VPRO simulations.

Non-reactive experiments are conducted to characterize the effect of heat loss to the reactor walls on the ignition delay. In these experiments, oxygen in the oxidizer is replaced entirely by nitrogen, to ensure that a similar specific heat ratio is maintained and similar heat loss conditions exist between the reactive case and the non-reactive case. The measured pressure trace from the non-reactive experiment is subsequently converted to a volume trace using the mixture specific heat ratio (as a function of temperature) and the adiabatic core relations. The volume trace is used to perform a simulation in CHEMKIN-Pro, and the calculated (non-reactive) pressure traces are found to match the experimental (non-reactive) pressure traces very well. This volume trace is then imposed on reactive simulations to account for the heat loss effect during the compression stroke and the post-compression event on autoignition. The deduced volume traces used in this study are available from the authors upon request, or on their website http://combdiaglab.engr.uconn.edu/database/18-rcm-database.

Temperature at TDC is taken as the reference temperature for reporting ignition delay data, and is obtained from the variable volume profile (VPRO) type simulations at the end of compression; this temperature is reported as the compressed temperature $(T_C)$. This approach requires the assumption of an adiabatic core of gases in the reaction chamber, which is facilitated on the present RCM by the creviced piston described previously. This general approach has been validated previously by Mittal and Sung.[28] Although the experimental pressure traces exhibit no apparent heat release during the compression stroke, calculations are nevertheless performed and compared with and without reaction steps for each kinetic mechanism to ensure no significant chemical heat release is contributing to the determination of the temperature at TDC. For those cases, the temperature profile during the compression stroke is shown to be the same whether or not reactions are included in the simulation.

**2.5 Definition of Ignition Delay**

The end of compression, when the piston reached TDC, is identified by the maximum of the pressure trace prior to the ignition point. The local maximum of the derivative of the pressure with respect to time, in the time after TDC, is defined as the point of ignition. The ignition delay is then defined as the time difference between the point of ignition and the end of compression. Figure 1 illustrates the definition of ignition delay used in this study, where $P(t)$ is the pressure trace and $P'(t)$ is the time derivative of the pressure trace. Also indicated on Figure 1 is the corresponding non-reactive pressure trace, which is obtained as described previously.

It can be seen that the pressure, and therefore temperature, decreases after the end of compression due to some heat loss to the reactor walls. Furthermore, as shown by Mittal et al.[32] there may be a small radical pool buildup in the reaction chamber during the compression stroke. The combination of these two effects will change the ignition delays relative to a corresponding adiabatic, constant-volume case whose simulated initial conditions match the TDC conditions in the reaction chamber. While the effect of heat loss generally tends to lengthen the ignition delay as compared to an adiabatic counterpart, under certain conditions the dominance of the effect of the radical pool buildup during the compression stroke can lead to a shorter ignition delay relative to a corresponding adiabatic, constant-volume case. Since these effects are very strongly fuel, equivalence ratio, temperature, and pressure dependent, this underlines the importance of conducting VPRO simulations in order to properly capture the combined effects.

## 3. Results and Discussion

### 3.1 Experimental Results

Figure 2 shows the ignition delays of the four isomers of butanol measured in the RCM, at compressed pressure of $P_C$ = 15 bar for stoichiometric mixture in air. The dashed line for each isomer is a least squares fit to the data. The vertical error bars are two standard deviations of the measurements of the ignition delay. The standard deviation is computed based on all the runs at a particular compressed temperature and pressure condition. A conservative estimate of the uncertainty in $T_C$ was calculated in our previous work to be approximately 0.7–1.7%.[16] Due to the similar nature of these experiments, and the similar properties of the fuels, this estimate is considered to be valid for this study as well.

Figure 2 demonstrates the differences in reactivity between the isomers for stoichiometric fuel/air mixtures at compressed pressure $P_C = 15$ bar. $n$-Butanol is clearly the most reactive, followed by $s$- and $i$-butanol, which have very similar reactivities in this temperature and pressure range. $t$-Butanol is the least reactive.

The order of reactivity found in the RCM at 15 bar agrees with the shock tube study at higher temperatures (approximately 1275−1667 K) and lower pressure (1.5 atm) by Stranic et al.,[17] but differs slightly from the studies of Moss et al.,[23] who measured ignition delays in a shock tube near 1.5 atm and between 1275−1400 K, and Veloo and Egolfopolous,[9] who measured atmospheric-pressure laminar flame speeds. In particular, Moss et al.[23] and Veloo and Egolfopolous[9] found distinct differences in reactivity between $s$- and $i$-butanol, but the present study and the study by Stranic et al.[17] found that they were nearly indistinguishable in terms of reactivity under the conditions investigated. In addition, Stranic et al.[17] noted some disagreement between their shock tube ignition data and the data of Moss et al.,[23] but their attempts to isolate the cause could not discern what the difference might be caused by.

Further, the order of the reactivity of the butanol isomers also shows complex temperature and pressure dependence. This is corroborated by the results shown in Figure 3. In Figure 3, the order of reactivity is different than in Figure 2, where the only variation between the plots is the compressed pressure; in Figure 3 the compressed pressure is $P_C = 30$ bar. Figure 3 shows *i*-butanol to be the least reactive, *s*-butanol to be less reactive than *t*-butanol (but similar), and *n*-butanol to be the most reactive. Interestingly, the results of the shock tube study by Stranic et al.[17] differ from those in the current study at higher pressure (despite the agreement at lower pressure). In their study, Stranic et al.[17] found *i*- and *n*-butanol to have similar reactivity near 43 atm. in the temperature range of 1020–1280 K, whereas in the present study we found *i*-butanol to be the least reactive of all four isomers at a pressure of 30 bar and over the temperature range (715–910 K) investigated.

The fact that *t*-butanol becomes relatively more reactive than *i*- and *s*-butanol as pressure increases is surprising at first glance, and the reasons are not immediately apparent. Closer examination of the pressure traces for each experiment gives one clue as to the cause of the increased reactivity. Figure 4 shows the pressure traces for the *t*-butanol experiments at 15 bar for stoichiometric mixtures in air. It is evident that there is some pre-ignition heat release, because the reactive pressure trace diverges from the non-reactive case prior to the ignition event. Of the other isomers of butanol, only *n*-butanol shows any visible heat release prior to the main ignition event at 15 bar.

Figure 5 shows the pressure traces for *t*-butanol experiments at 30 bar for stoichiometric mixtures in air. The effect of pre-ignition heat release is even more striking in this figure, with substantial changes in the slope of the pressure trace during the reactive runs. Comparing to the pressure traces of the other isomers once again shows that the magnitude of the pre-ignition heat release for *t*-butanol is much greater. Despite the appearance of early pressure rise, which is typically indicative of two-stage ignition and low temperature chain branching, we do not find a negative temperature coefficient region in terms of the ignition delay response for any *t*-butanol experiments. Therefore, we adopt the phrase "pre-ignition heat release" rather than "two-stage ignition" in this work.

In an effort to understand the reactions causing the pre-ignition heat release, further experiments are conducted for *t*-butanol at $P_C$ = 30 bar, for equivalence ratios of 0.5 and 2.0 in air. Figure 6 shows Arrhenius plots of the ignition delays for the three equivalence ratios. As with the previous *n*-butanol experiments at 15 bar,[16] $\phi$ = 0.5 is the least reactive and $\phi$ = 2.0 is the most reactive. The slopes are similar, indicating that the overall activation energies are similar for the conditions investigated.

A more interesting comparison is of the pressure traces of the three equivalence ratios. It is clear from Figures 5, 7, and 8 that there are qualitative differences in the pre-ignition heat release between the three equivalence ratios. This is most likely due to the effect of the increased (reduced) fuel mole fraction in the $\phi$ = 2.0 ($\phi$ = 0.5) case, since the mole fraction of fuel is changed by +93% (-49%) compared to the $\phi$ = 1.0 case, while the mole fraction of oxygen changes by only -3% (+2%) compared to the $\phi$ = 1.0 case, as shown in Table 1. Therefore, it appears that the qualitative change in pre-ignition behavior is due to the change of fuel mole fraction, where higher fuel loading promotes pre-ignition heat release.

## 3.2 Simulation Results

Simulations are performed with the kinetic mechanism from Sarathy et al.[18] and a recent mechanism discussed in Hansen et al.[19] and Merchant et al.[20] that is denoted as the MIT mechanism hereafter. Other recent mechanisms, such as the mechanism from Frassoldati et al.,[33] do not include low temperature chemistry and are therefore unable to reproduce the low-temperature ignition delays measured in this study. The study by Sarathy et al.[18] validated their model for a wide set of the existing experimental data. In terms of ignition delays, this included the data from the study of Stranic et al.[17] up to 48 atm, our previous study on *n*-butanol,[16] and the data being published in this study at 15 bar. Importantly, the mechanism of Sarathy et al.[18] was validated only for the 15 bar RCM data for all four isomers, but not the 30 bar data also being published here. The MIT mechanism[19,20] was validated for *i*-butanol experiments, including pyrolysis and low pressure premixed flames; although the model includes all four isomers of butanol as reactants, it has not been optimized for any of the isomers except *i*-butanol.

Figures 9 and 10 show comparison of the VPRO simulations with the experimental data using the mechanism of Sarathy et al.[18] As Sarathy et al.[18] showed in their work (and as we show here in Figure 9), they found good agreement of the model predictions with the present RCM data at 15 bar. At $P_C$ = 30 bar (Figure 10), similar degree of agreement is found for *t*-butanol and *s*-butanol compared to $P_C$ = 15 bar, although the *s*-butanol results are under-predicted at high temperature and over-predicted at low temperature. While the model of Sarathy et al.[18] is able to well capture the overall activation energy of *i*-butanol, it under-predicts the experimental data by about a factor of 2–3. The *n*-butanol data are over-predicted by about a factor of 1.5. Nevertheless, this agreement is quite good, especially considering that the model is not validated for these conditions.

VPRO simulations for *n*- and *s*-butanol (and also some conditions for *t*-butanol) using the MIT mechanism[19,20] do not ignite during the duration of the simulations (the same as the experimental duration), and therefore no simulations are shown for these fuels. In Figure 11, VPRO simulations at 15 and 30 bar using both mechanisms are shown for *i*-butanol. It is seen that the mechanism from Sarathy et al.[18] is in better agreement at 15 bar. However, at 30 bar the MIT mechanism over-predicts the ignition delay (as at 15 bar), while the Sarathy et al.[18] mechanism under-predicts the ignition delay. The reason for these diverging predictions will be explored and discussed below.

The agreement of the mechanism by Sarathy et al.[18] with the off-stoichiometric mixtures of *t*-butanol is also quite good, as shown in Figure 12. Figures 13–15 show more detailed comparisons of the simulated pressure traces and the experimental results, for similar temperatures at the three equivalence ratios, respectively. Clearly, the simulations also exhibit some pre-ignition heat release. In general, the simulations qualitatively predict the pre-ignition heat release behavior at all three equivalence ratios. The $\phi = 0.5$ case has the least heat release and the $\phi = 2.0$ case has the most. Although the simulations are unable to match the heat release behavior quantitatively, they match the experimental ignition delays quite well. Considering the model is not validated for this temperature, pressure, and equivalence ratio regime, the mismatch of the pre-ignition behavior may not be of critical importance, depending on the application.

**3.3 Discussion**

The relatively good agreement of the mechanism of Sarathy et al.[18] with the experimental data, even for conditions at which the mechanism has not been validated, suggests that using the mechanism to further interpret our experimental data is not a facile exercise. In particular, Figures 16–19 show the initial steps of the fuel breakdown process for each isomer. The percentages listed are the percent of the reactant that is consumed to produce the product shown, by all the reactions that can produce that product from the reactant (except where one particular reaction is noted). These numbers are determined by integrating the rate of production or consumption of each species by each reaction up to the point of 20% fuel consumption, and normalizing each reaction by the total produced or consumed of each species up to that point. The 20% fuel consumption point is chosen because it is before small molecule chemistry takes over to drive the ignition, and it has been used previously.[16,18] The rates of production are taken from a CONV simulation, with initial conditions 750 K and 15 bar as well as 750 K and 30 bar. These conditions are representative of typical conditions after compression in the present RCM experiments. The plain text percentages on top of the arrows are the 15 bar case and the bold numbers underneath are for the 30 bar case.

In the following discussion, carbon-centered radicals are labeled according to their distance from the hydroxyl moiety in the fuel molecule. Therefore, the $\alpha$ carbon is the closest to the hydroxyl, followed by $\beta$, $\gamma$, and $\delta$ carbons. Not all of the butanols have all of the types of carbons listed here, due to varying chain lengths. For instance, *t*-butanol has one $\alpha$ carbon, three $\beta$ carbons, and no $\gamma$ or $\delta$ carbons.

As expected at the relatively low temperature of this analysis, H-abstraction reactions dominate over unimolecular decomposition for all four isomers. It is also expected that *n*-, *s*-, and *i*-butanol react primarily to their respective $\alpha$-hydroxybutyl radicals, since the $\alpha$ C-H bond has the lowest energy.[18] Due to its unique structure, *t*-butanol does not have an $\alpha$-hydroxybutyl radical that can be formed by H-abstraction, so *t*-butanol is primarily consumed to form the $\beta$-hydroxybutyl radical, because the O-H bond energy is much higher than $\beta$ C-H bond energies.

The unique structure of *t*-butanol continues to affect the second level of reactions as well. In the temperature and pressure regime investigated, *t*-butanol tends to add to molecular oxygen at the carbon radical site, forming a hydroxybutylperoxy ($RO_2$) species. That this pathway is dominant is due to the fact that *t*-butanol has no $\alpha$-hydroxybutyl radical. For the other three butanol isomers that do have an $\alpha$-hydroxybutyl radical, the second level of reactions primarily produces an aldehyde + $HO_2$ by direct reaction – no hydroxybutylperoxy adduct is formed in this reaction, and there is no possibility for typical hydrocarbon low-temperature chain branching. Therefore, it is hypothesized that the pre-ignition heat release seen in *t*-butanol is caused by the oxygen addition to the fuel radical to form $\beta$-hydroxybutylperoxy, which is an exothermic reaction.

Figure 20 shows the total cumulative heat release of each isomer and the cumulative heat release of an important reaction for each of the isomers (inset), from a CONV simulation with initial conditions of 750 K and 30 bar; analysis of 15 bar results is substantially similar. The cumulative heat release in the inset is found by integrating the heat release by each reaction with respect to time, while the reactions shown are the respective reactions that have released the most heat up to the 20% fuel consumption point for each isomer. The abscissa of the plot is the fuel conversion, in percent. This choice of x-axis allows a fair comparison of the heat release, because the ignition delays of each isomer are markedly different, so comparing the heat release with a time axis is more difficult. In Figure 20, exothermicity is represented by positive quantities.

In Figure 20, it is clear that *t*-butanol has higher heat release at low fuel consumption (during the induction period) than the other three isomers. In addition, the primary heat release reaction for *t*-butanol has created much more heat than the primary reactions of the other three isomers. As the reactions proceed, and the temperature increases, the reverse reaction in the *t*-butanol case becomes more important, and the heat release contribution of this oxygen-addition reaction levels off. The dominance of this reaction at early times is unique to *t*-butanol ignition, and appears to be driving the pre-ignition heat release.

Other researchers have also undertaken studies of the low to intermediate temperature combustion of *t*-butanol. Lefkowitz et al.[15] performed a study in the Variable Pressure Flow Reactor (VPFR) at Princeton University on the oxidation of *t*-butanol over the temperature range from 680−950 K, at 12.5 atm and stoichiometric mixture conditions. It is interesting to note that they found no evidence of traditional hydrocarbon low temperature chemistry. They did, however, find significant quantities of acetone, peaking at approximately 800 K. Lefkowitz et al.[15] concluded that the primary pathways of acetone formation are tautomerization of propen-2-ol and β-scission of the alkoxy radical, based on an analysis of the mechanism from Grana et al.[13] Both of these pathways are dependent on unimolecular decomposition of the hydroxybutyl radicals. However, this mechanism has only been validated for flame studies; indeed, an updated version of this model (by Frassoldati et al.[33]) is unable to predict the low-temperature ignition delays measured in this study and hence is not considered for analysis.

In contrast to the study of Lefkowitz et al.,[15] path analysis of the mechanism by Sarathy et al.[18] shows that unimolecular decomposition of the hydroxybutyl radicals is not the most important pathway; as mentioned earlier, the most important pathway is the formation of β-hydroxybutylperoxy. Further analysis shows that the primary pathway of reaction of the *t*-butanol β-hydroxybutylperoxy species is through the Waddington mechanism. The Waddington mechanism has been shown experimentally to be an important pathway for β-hydroxypentylperoxy radicals in the low temperature combustion of *i*-pentanol,[34] as well as the β-hydroxybutylperoxy radicals of *i*- and *t*-butanol.[35] *t*-Butanol only produces β-hydroxybutyl radicals, and one of the products of the Waddington pathway in *t*-butanol is acetone (the others are formaldehyde and hydroxyl radical); over 88% of the acetone produced up to the 20% fuel consumption point is produced by the Waddington reaction. The study in the VPFR thus provides further evidence of the importance of low-temperature hydroxybutylperoxy chemistry in *t*-butanol, although it is not traditional hydrocarbon low-temperature chemistry.

Up to this point, the discussion has focused mainly on the importance of hydroxybutylperoxy chemistry in *t*-butanol. Nevertheless, the chemistry of the hydroxybutylperoxy species is important in the combustion of the other isomers of butanol as well. Using the high pressure shock tube at RWTH Aachen University, Vranckx et al.[26] showed the importance of peroxy chemistry pathways in the autoignition of *n*-butanol. By adding a lumped peroxy model to an existing kinetic model for *n*-butanol combustion, they were able to substantially improve agreement of the model with their experiments at high pressure and low temperature. In their mechanism, Sarathy et al.[18] included a semi-detailed peroxy chemistry model for all the isomers of butanol. In fact, one of the main differences between the mechanism from Sarathy et al.[18] and the MIT mechanism[19,20] is their respective treatment of the peroxy mechanism. Specifically, oxygen-addition chemistry is not included in the MIT mechanism for *i*-butanol.[19,20] In addition, the radical that primarily controls *i*-butanol decomposition is hydroxyl (OH) in the mechanism of Sarathy et al.[18] (generated by the peroxy chemistry sub-mechanism), but is hydroperoxyl ($HO_2$) in the MIT mechanism[19,20] (generated from the direct $\alpha$-hydroxybutyl + $O_2$ = $HO_2$ + aldehyde formation pathway).

In their work, Sarathy et al.[18] used the reaction rates computed by Da Silva et al.[36] for the hydroxyethyl system (i.e. ethanol as the parent fuel) to determine the rate of direct reaction of $\alpha$-hydroxybutyl and oxygen to form aldehyde and $HO_2$, and then set the rate of oxygen addition to the $\alpha$-hydroxybutyl radical (to form $\alpha$-hydroxybutylperoxy) so that the total rate was less than the collisional limit. The rates of oxygen addition for the other radicals were prescribed depending on the type of carbon (primary, secondary, or tertiary) based on studies of butane and *i*-octane.[18] Based on the well-known importance of hydroxyl in driving the reactivity of combustion systems, and the sources of the estimates for the reaction rates of oxygen addition to hydroxybutyl (i.e. the entry to the pathway that controls the rate of hydroxyl formation), it can be hypothesized that the rates of hydroxybutylperoxy formation are overestimated in the mechanism of Sarathy et al.,[18] as the simulated results under-predict the experimental data of *i*-butanol.

This hypothesis is supported by the results shown in Figure 21, which shows the linear brute force sensitivity of the ignition delay ($\tau$) of *i*-butanol to changes in the A-factor of the rate coefficient, using the mechanism from Sarathy et al.[18] The percent sensitivity is defined as the difference between the ignition delay when the A-factor of each reaction is halved and the nominal ignition delay, normalized by the nominal ignition delay, as shown below:

$$\% \text{ Sensitivity} = \frac{\tau(0.5A_i) - \tau(A_i)}{\tau(A_i)} \times 100\%.$$

Therefore, negative sensitivity means that halving the A-factor of a reaction decreases the ignition delay, and positive sensitivity indicates the ignition delay increases. These results are for CONV simulations with initial conditions of 750 K and 30 bar as well as 1200 K and 30 bar.

The most sensitive reaction at the lower temperature is the initiation reaction of the fuel with hydroperoxyl radical to form the primary fuel radical and the second most sensitive reaction is the addition of oxygen to the primary radical. Both of these reactions have positive sensitivities, indicating that reducing the rate of these reactions would increase the ignition delay (increasing the computed ignition delay will improve the agreement of the simulations relative to the experiments in this case). It is apparent, then, that reducing the amount of fuel propagating into the low temperature chain branching pathway of oxygen addition to the primary $\alpha$-radical improves the simulated results. Interestingly, the *i*-butanol system is not sensitive to the rates of oxygen addition to the hydroxybutyl radicals other than the $\alpha$-radical. At the higher temperature of 1200 K, there is little sensitivity on the ignition delay by changing the rate of the oxygen-addition reaction, demonstrating its lack of influence at higher temperatures.

As a final comparison, we have modified this pathway in the mechanism from Sarathy et al.,[18] so that the rate of oxygen addition to the primary fuel radical is arbitrarily set to zero; that is, the rate of the reaction ic4h8oh-1+$O_2$⇔ic4h8oh-1$O_2$ is set to zero by zeroing the A-factor, while the rates of the other oxygen addition reactions were unchanged. This unphysical situation substantially changes the results of simulations for *i*-butanol − removing this pathway in the mechanism from Sarathy et al.[18] brings the simulations into close agreement with the ignition delay results from the MIT mechanism,[19,20] which does not consider this reaction for *i*-butanol. Since the other oxygen addition reactions were unchanged, it is apparent that the addition of oxygen to $\alpha$-hydroxybutyl is one of the controlling reactions for the high-pressure, low-temperature ignition of *i*-butanol using the mechanism of Sarathy et al.[18] It is therefore concluded that a detailed examination of the rates of direct formation of aldehyde+$HO_2$ and oxygen addition to the $\alpha$-hydroxybutyl radical are required to better predict the low-temperature ignition behavior of *i*-butanol. Furthermore, based on the other results of this study, a detailed analysis of the oxygen addition reactions to all the isomers of butanol is probably warranted.

## 4. Conclusions

In this work, ignition delays for all four isomers of butanol in stoichiometric mixture with air have been presented over the low to intermediate temperature range, and at two compressed pressures of 15 and 30 bar. The order of reactivity of the isomers goes *n*-butanol>*s*-butanol≈*i*-butanol>*t*-butanol at the lower pressure, but changes to *n*-butanol>*t*-butanol>*s*-butanol>*i*-butanol at the higher pressure. This unexpected result is partially explained by the fact that there is substantial pre-ignition heat release present for *t*-butanol. To help understand the nature of the pre-ignition heat release of *t*-butanol, studies at off-stoichiometric conditions, $\phi = 0.5$ and $\phi = 2.0$ in air, are also conducted.

Comparisons of the experimentally measured ignition delays with two kinetic mechanisms show good agreement for certain isomers, but relatively poorer agreement for others. The kinetic mechanism of Sarathy et al.[18] is used to further elucidate the chemical processes controlling the autoignition of these butanol isomers. Pathway analysis of the fuel decomposition shows that *n*-, *s*-, and *i*-butanol primarily form $\alpha$-hydroxybutyl radicals, because the proximity of the $\alpha$-carbon to the hydroxyl group reduces the C-H bond energy. The $\alpha$-hydroxybutyl radicals tend to form an aldehyde plus $HO_2$ directly, without forming a hydroxybutylperoxy complex. However, due to its unique structure, *t*-butanol can only form $\beta$-radicals; these radicals do not have the tendency to react with oxygen to directly form $HO_2$ and an aldehyde. Rather, *t*-butanol preferentially adds oxygen to the fuel radical site. It is hypothesized that this reaction, $O_2$ addition to form hydroxybutylperoxy, causes the pre-ignition heat release in *t*-butanol and leads to a chain propagation pathway through the Waddington mechanism. The fact that this oxygen-addition reaction is preferred is unique to *t*-butanol, although a detailed understanding of the peroxy chemistry of alcohols is still of vital importance to the other butanol isomers. This is further demonstrated in this work for the case of *i*-butanol, where the ignition delay is shown to be quite sensitive to both the rate of primary fuel radical formation and to the rate of oxygen addition to the primary fuel radical. It is also noted that *n*-butanol autoignition was shown to be quite sensitive to peroxy chemistry in the study of Vranckx et al.[26]

All together, these analyses show the importance of the peroxy chemistry pathways in the autoignition of the butanols. Further experimental studies, such as species profiles from the low temperature ignition of the butanol isomers, could help reduce uncertainty in the pathways of fuel breakdown. Finally, further understanding of the rates of the peroxy pathways is important and therefore further theoretical and quantum chemical studies are warranted.

**Acknowledgements**

The authors acknowledge support from the Combustion Energy Frontier Research Center, an Energy Frontier Research Center funded by the U.S. Department of Energy, Office of Science, Office of Basic Energy Sciences, under Award Number DE-SC0001198. The authors gratefully acknowledge Dr. William Green and Shamel Merchant of the Massachusetts Institute of Technology for providing their mechanism prior to publication, and for useful discussions; and Dr. Mani Sarathy of King Abdullah University of Science and Technology for very helpful discussions.

**Supporting Information Available**

Ignition delay measurements in the Rapid Compression Machine. This material is available free of charge via the Internet at http://pubs.acs.org.

**List of Tables**

**List of Figures**

Table 1: Experimental Conditions and Reactant Purities

| Reactant (Purity) | | | | | Equivalence Ratio | Compressed Pressure |
|---|---|---|---|---|---|---|
| s-butanol (99.99%) | i-butanol (99.99%) | t-butanol (99.99%) | O$_2$ (99.999%) | N$_2$ (99.995%) | $\phi$ | $P_C$ (bar) |
| Mole Percentage | | | | | | |
| 3.38 | | | 20.30 | 76.32 | 1.0 | 15 |
| 3.38 | | | 20.30 | 76.32 | 1.0 | 30 |
| | 3.38 | | 20.30 | 76.32 | 1.0 | 15 |
| | 3.38 | | 20.30 | 76.32 | 1.0 | 30 |
| | | 3.38 | 20.30 | 76.32 | 1.0 | 15 |
| | | 3.38 | 20.30 | 76.32 | 1.0 | 30 |
| | | 1.72 | 20.65 | 77.63 | 0.5 | 30 |
| | | 6.54 | 19.63 | 73.83 | 2.0 | 30 |

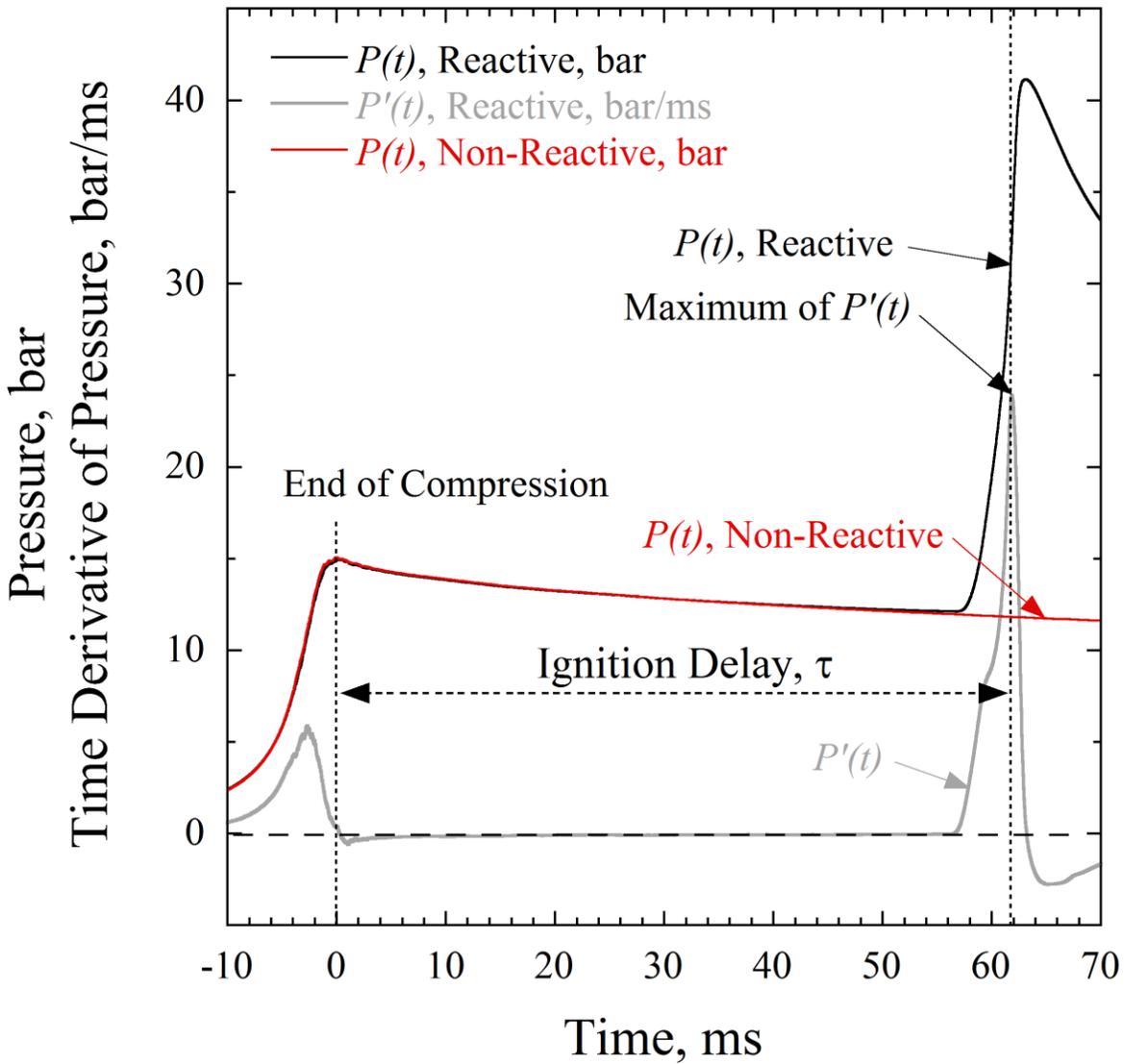

**Figure 1.** Definition of ignition delay used in this study. *P(t)* is the pressure as a function of time and *P'(t)* is the time derivative of the pressure, as a function of time. The corresponding non-reactive experiment is described in Section 2.4.

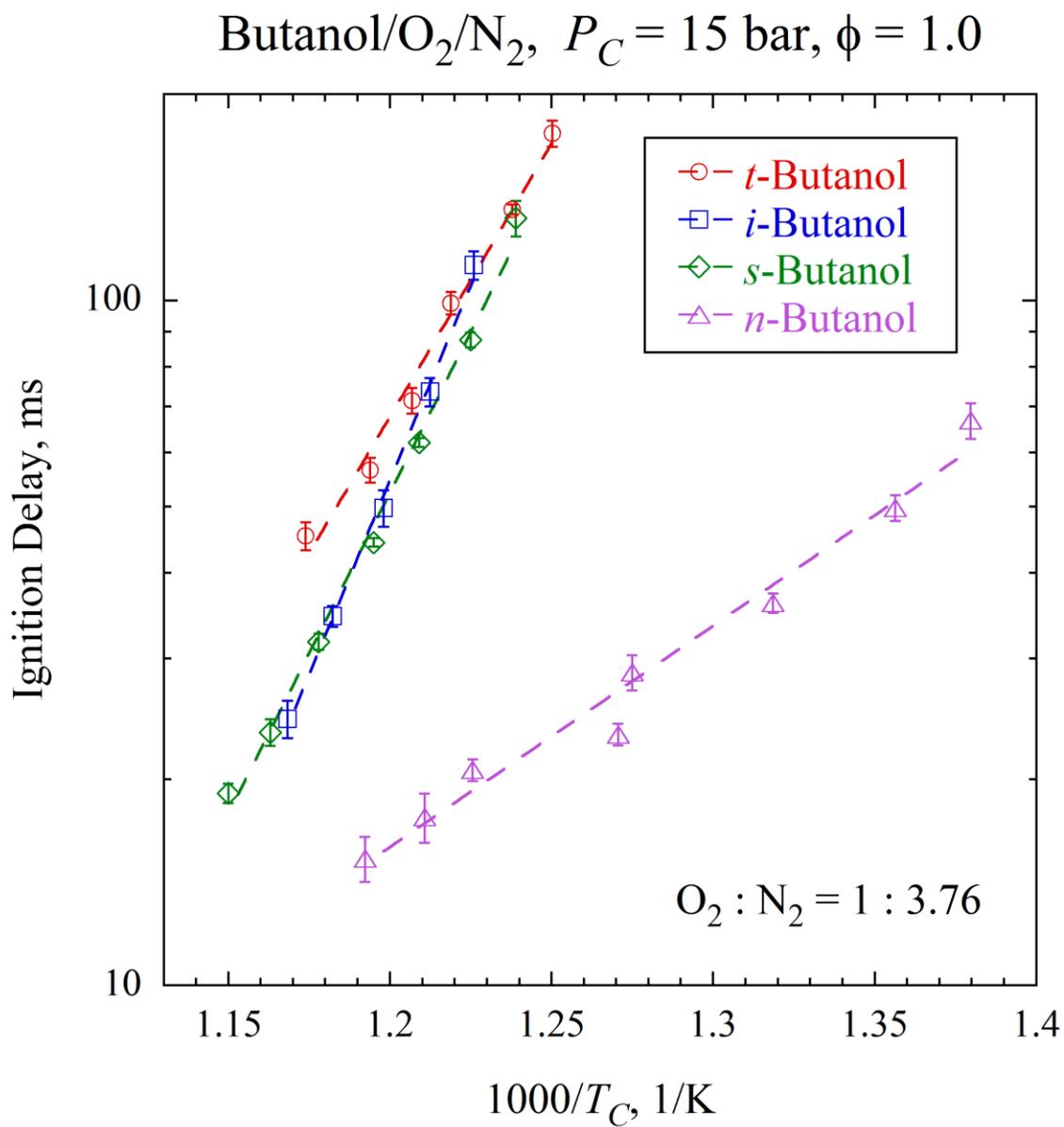

**Figure 2.** Ignition delays of the four isomers of butanol at compressed pressure $P_C = 15$ bar. Dashed lines are least squares fits to the data.

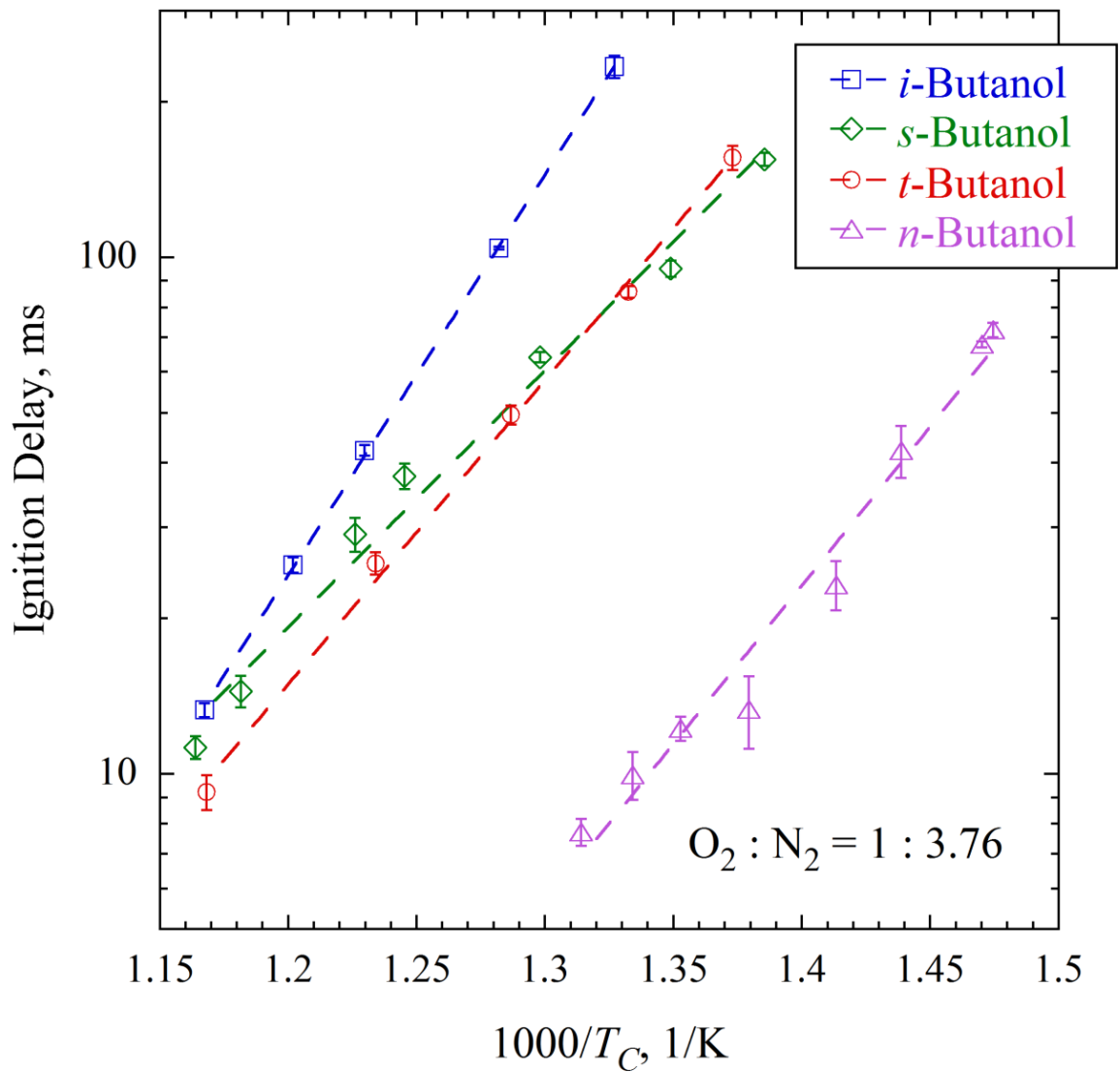

**Figure 3.** Ignition delays of the four isomers of butanol at compressed pressure $P_C$ = 30 bar. Dashed lines are least squares fits to the data.

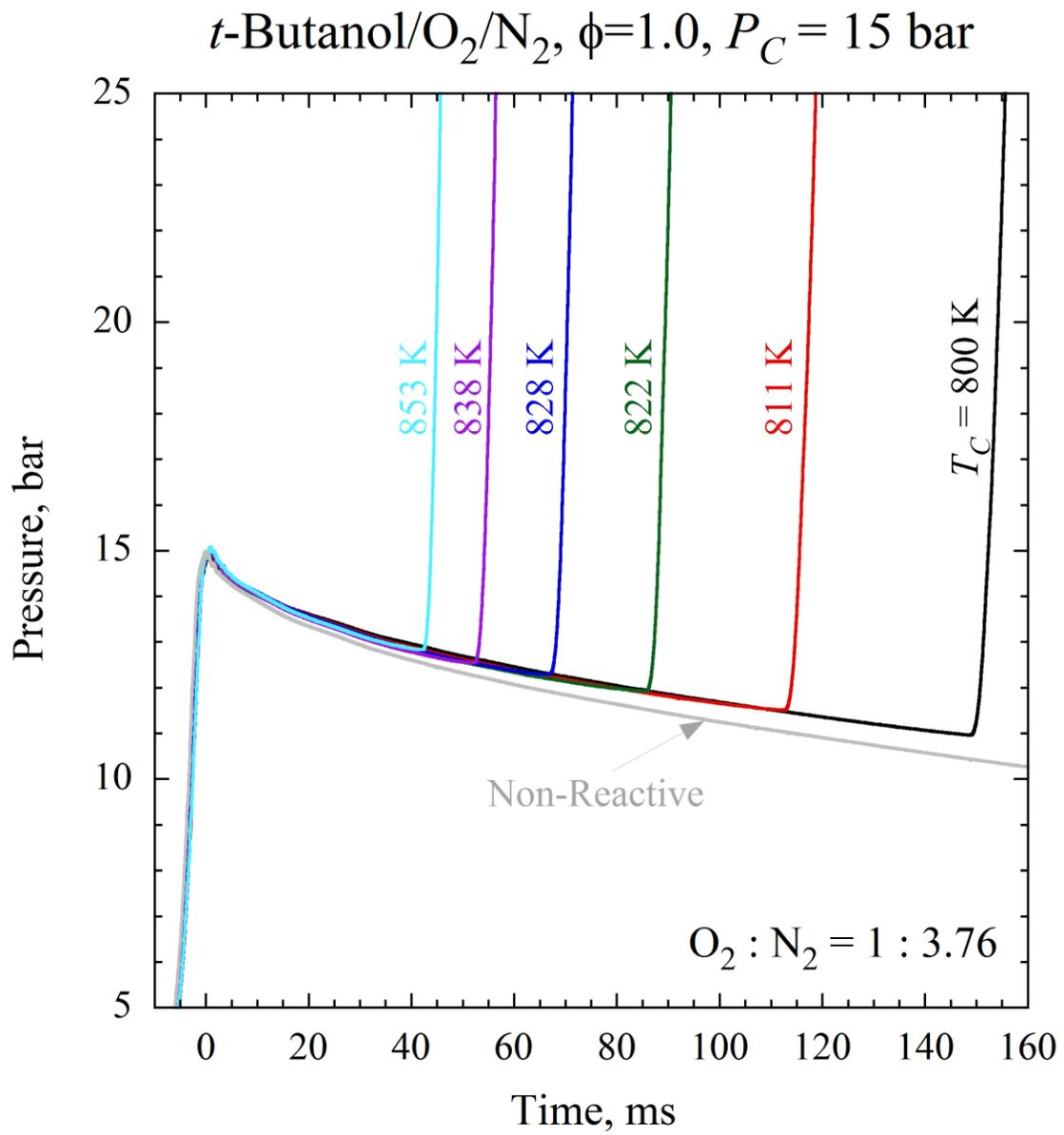

**Figure 4.** Pressure traces of the 15 bar *t*-butanol experiments, in stoichiometric air.

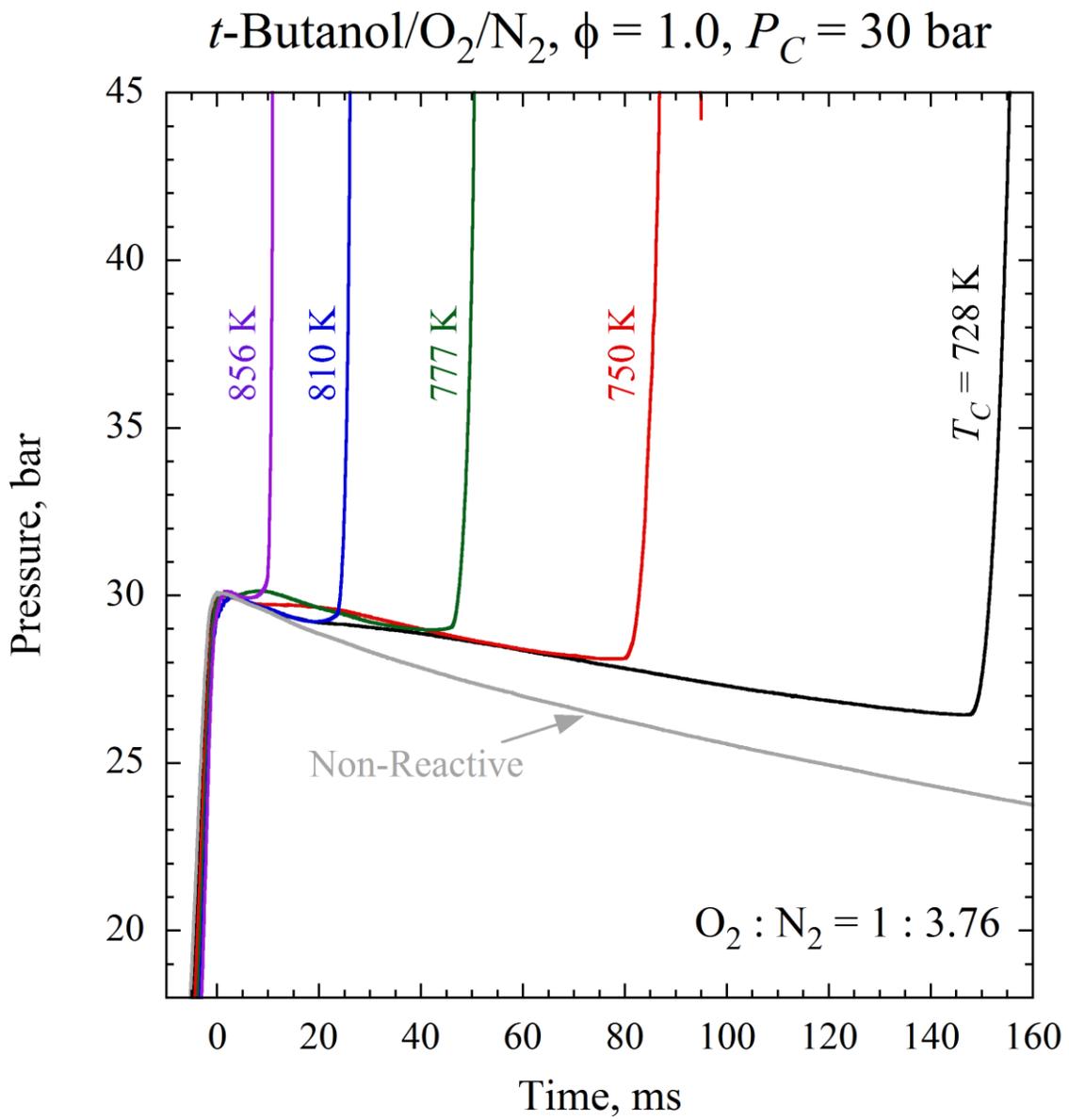

**Figure 5.** Pressure traces of the 30 bar *t*-butanol experiments, in stoichiometric air.

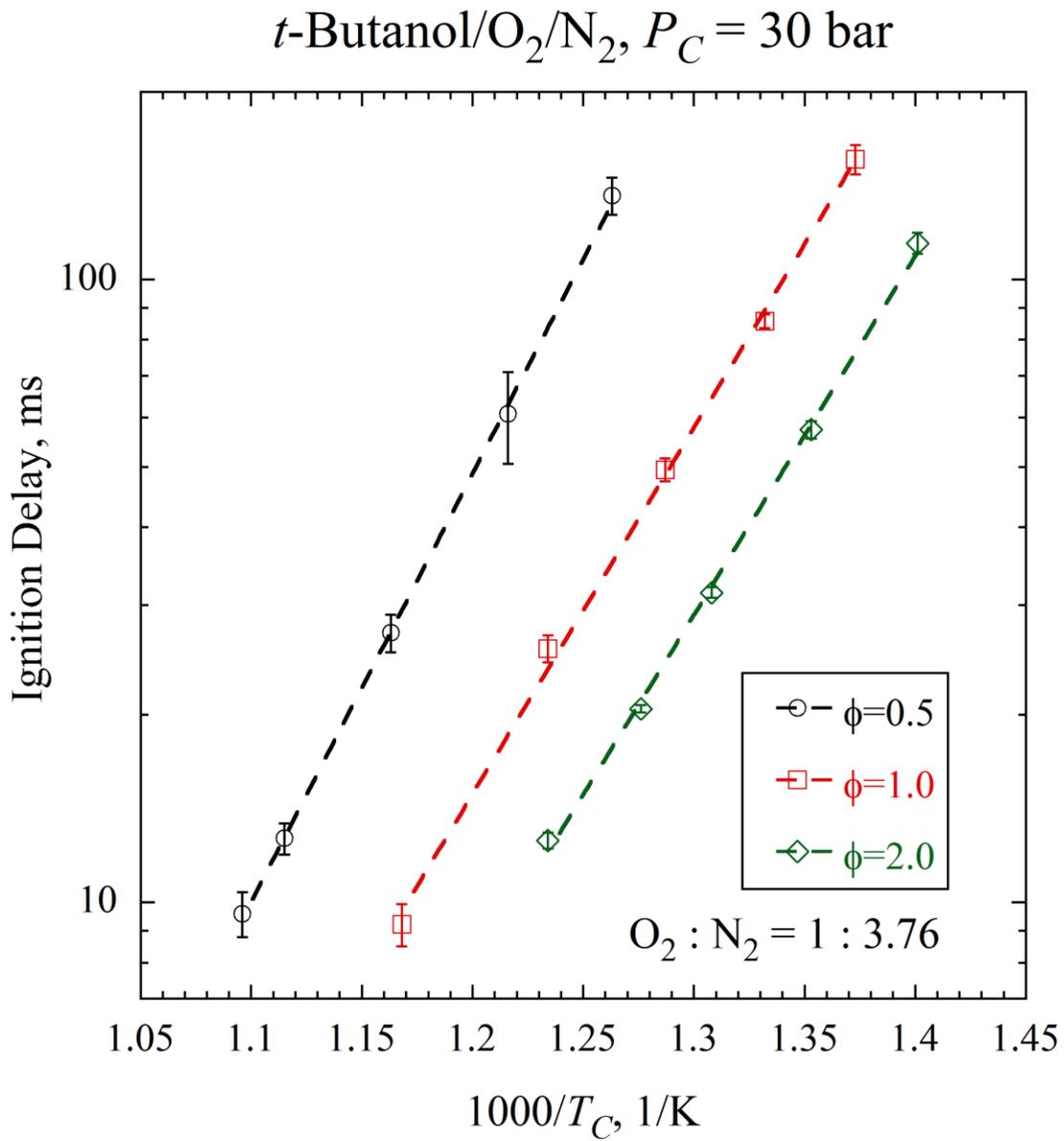

**Figure 6.** Ignition delays of three equivalence ratios of *t*-butanol in air, for $P_C = 30$ bar. Lines represent least squares fits to the data.

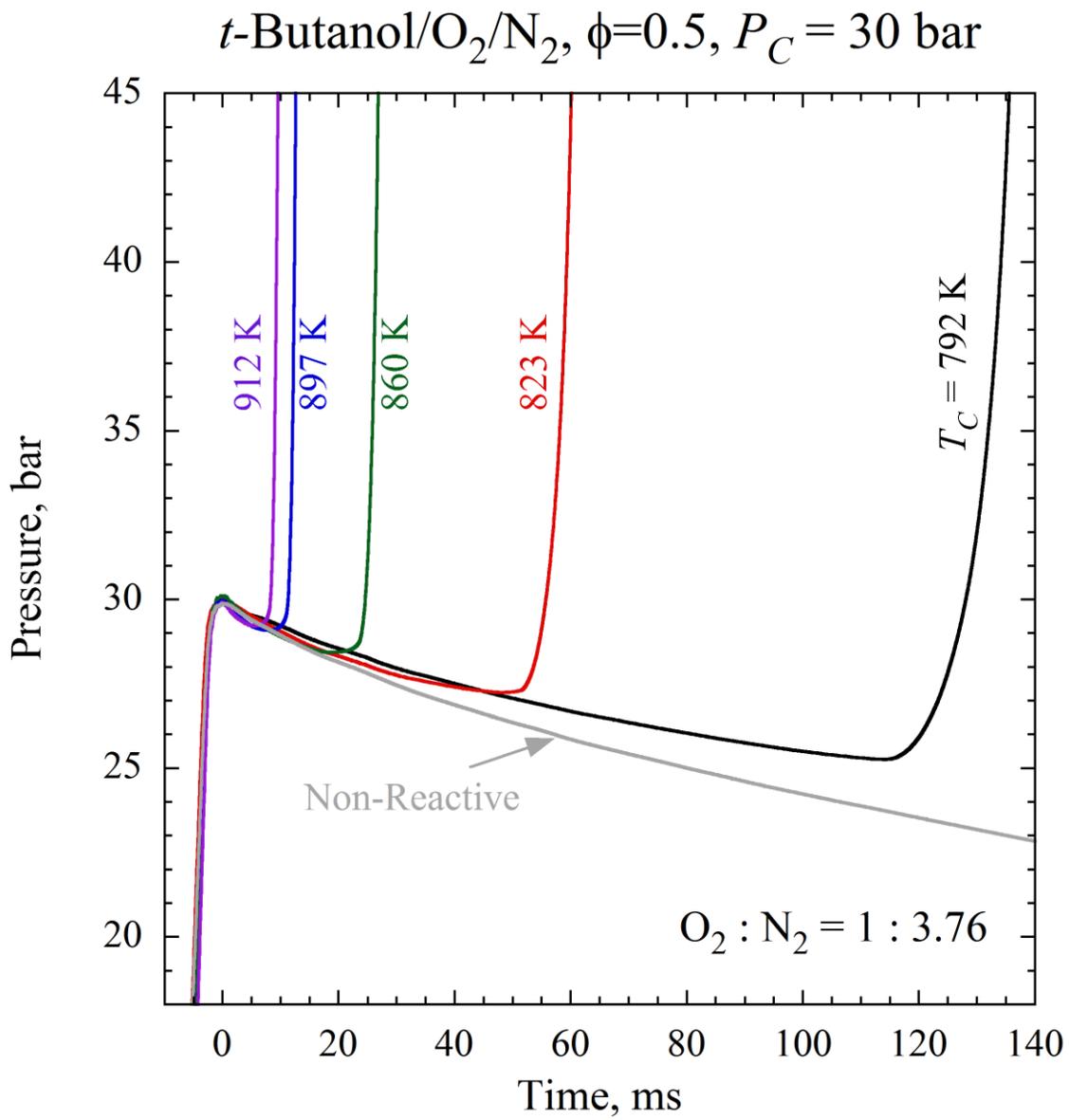

**Figure 7.** Pressure traces of the 30 bar *t*-butanol experiments, $\phi = 0.5$ in air.

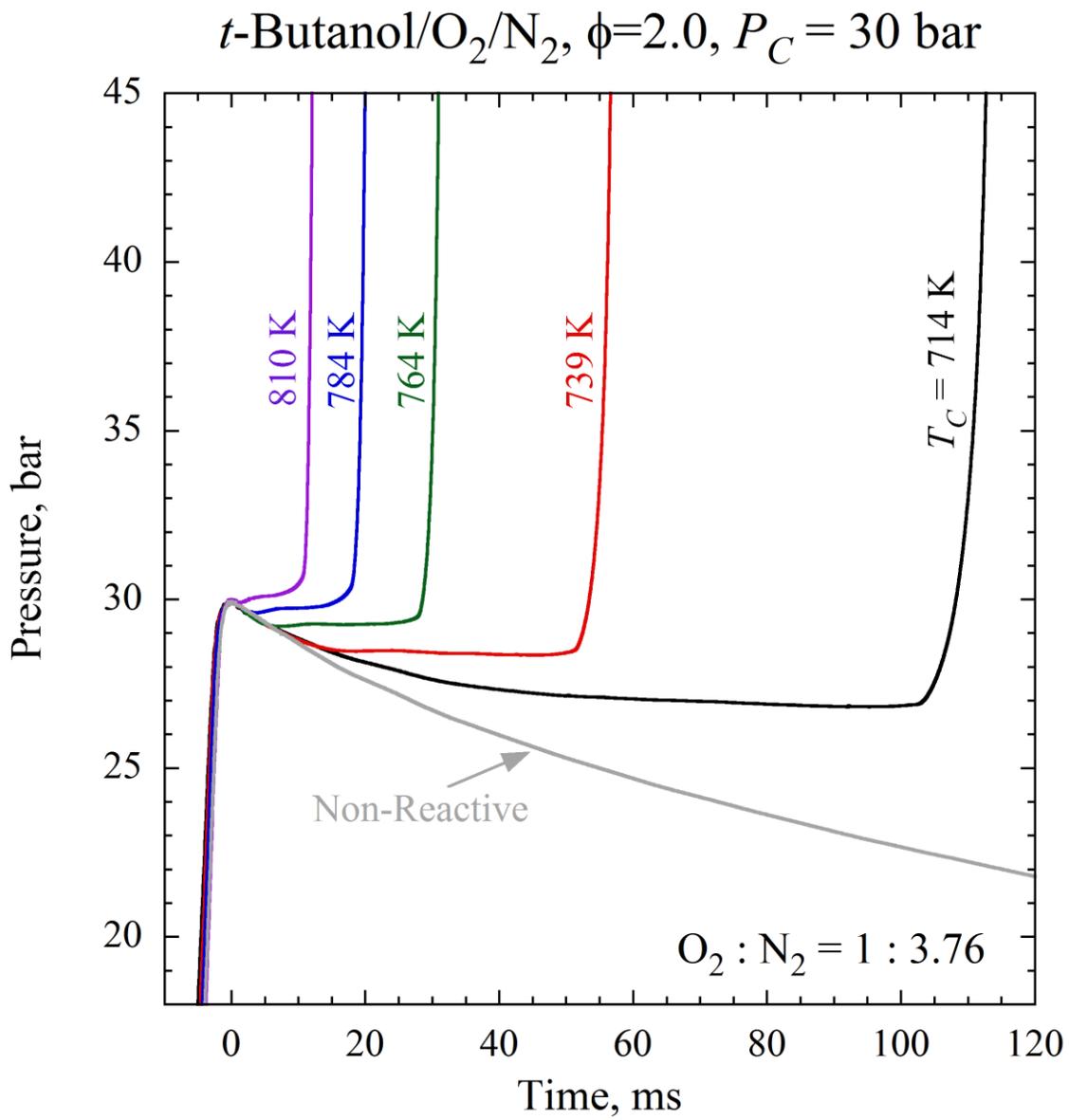

**Figure 8.** Pressure traces of the 30 bar *t*-butanol experiments, $\phi = 2.0$ in air.

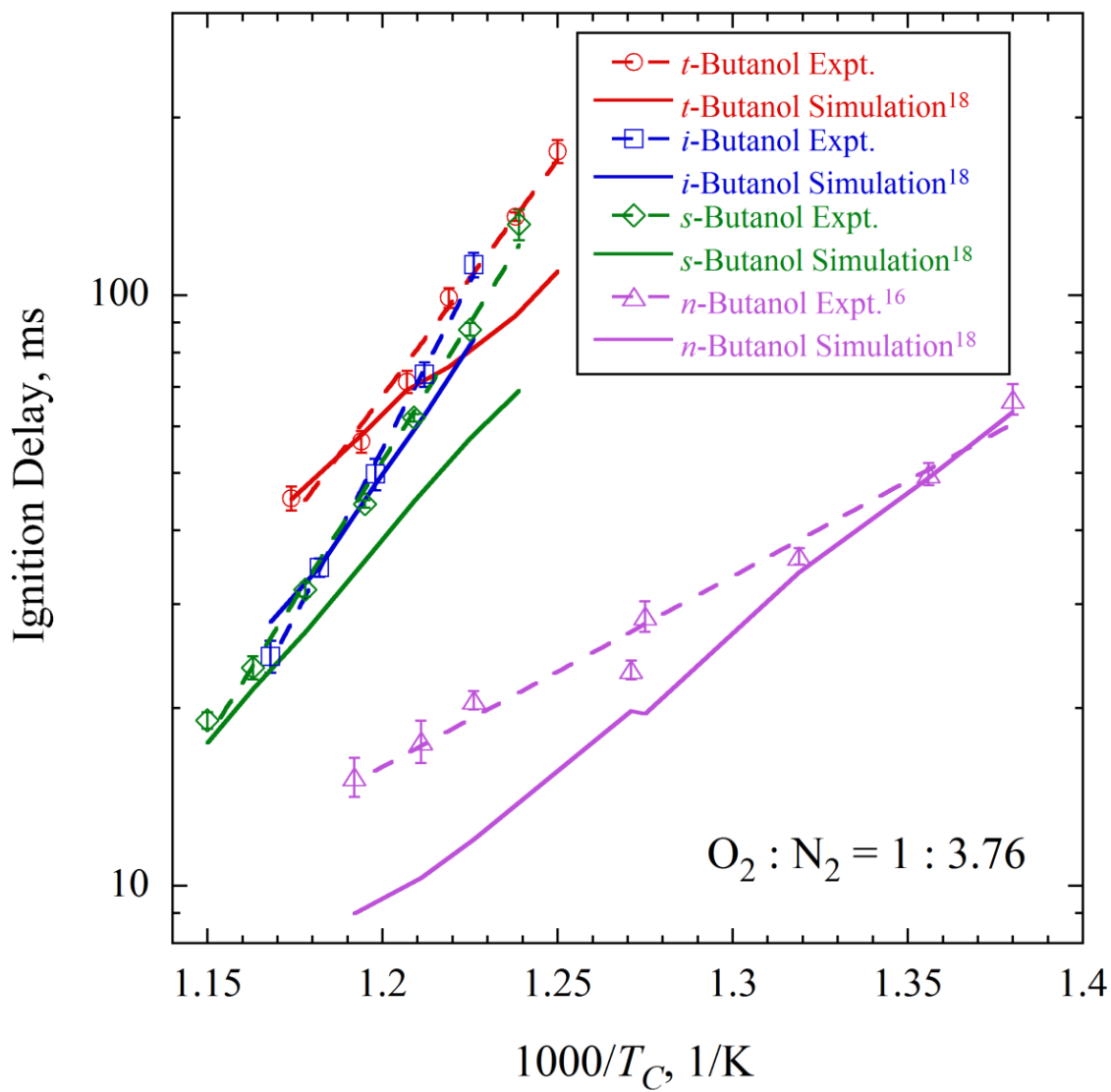

**Figure 9.** $P_C = 15$ bar, stoichiometric mixtures in air. Comparison of VPRO simulations using the kinetic mechanism of Sarathy et al.[18] with experimental ignition delay measurements.

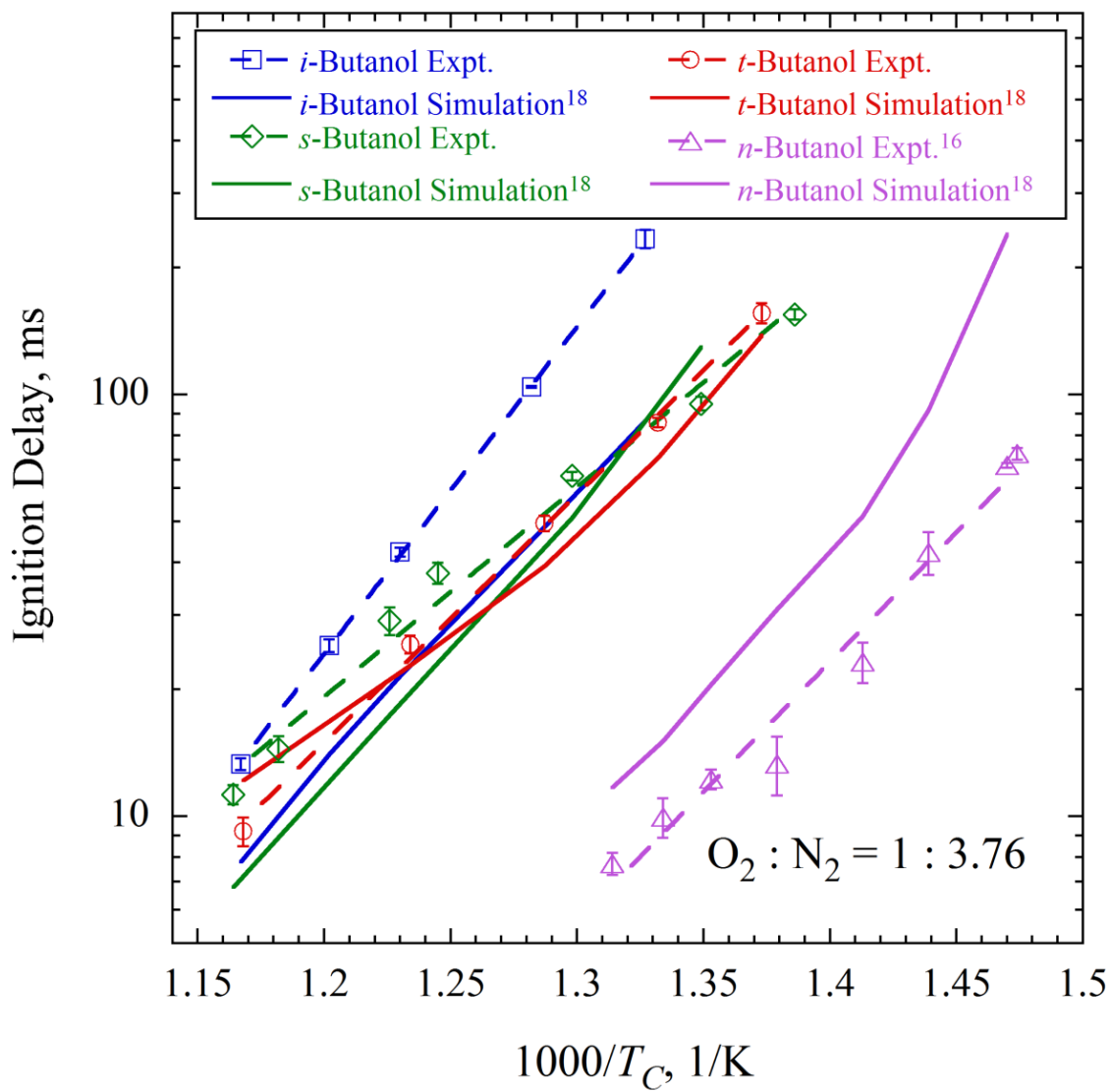

**Figure 10.** $P_C = 30$ bar, stoichiometric mixtures in air. Comparison of VPRO simulations using the kinetic mechanism of Sarathy et al.[18] with experimental ignition delay measurements.

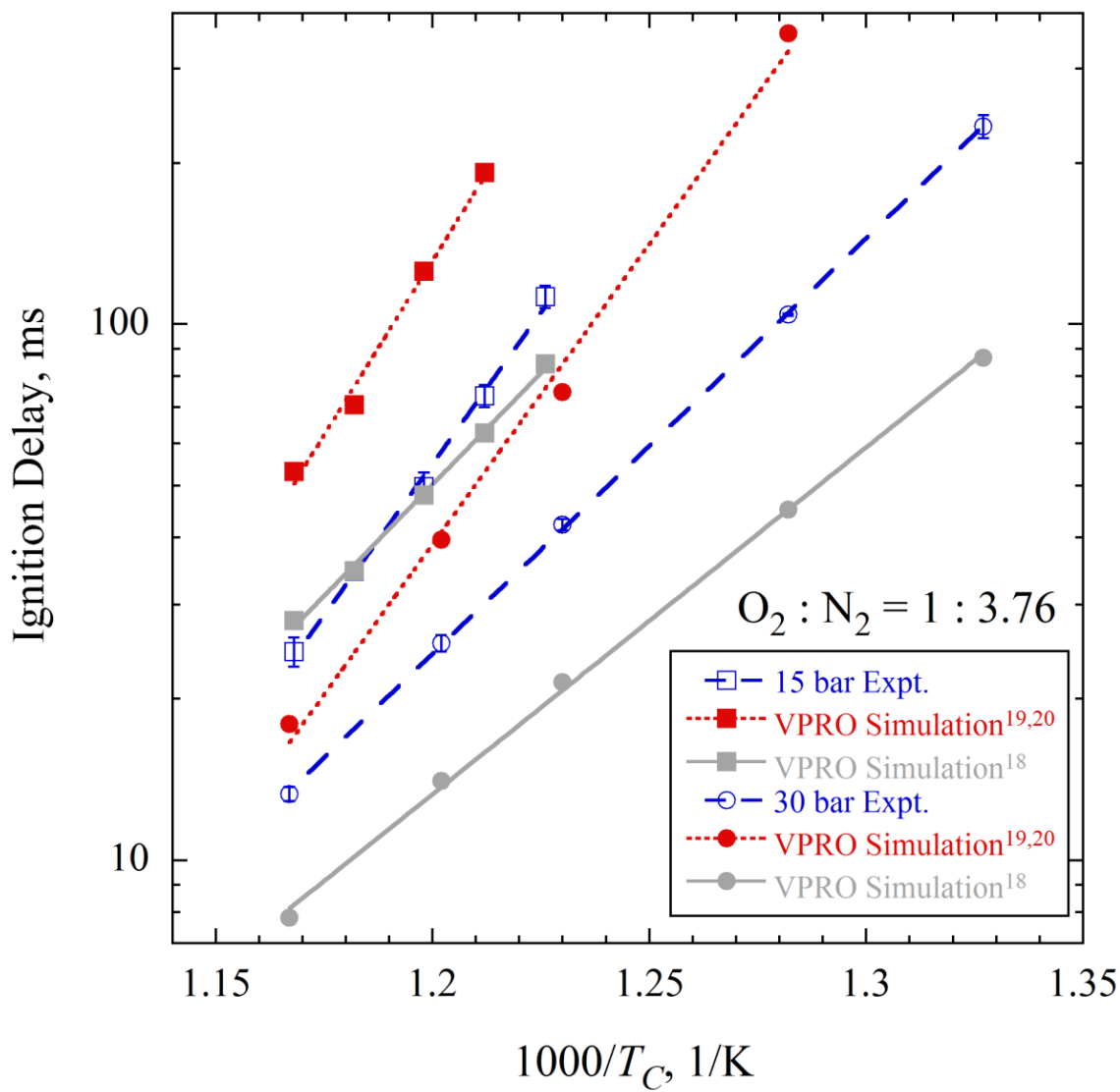

**Figure 11.** Comparison of VPRO simulations using the kinetic mechanism of Sarathy et al.[18] (solid lines) and the MIT mechanism[19,20] (dotted lines) with experimental ignition delay results (dashed lines) for stoichiometric mixtures of $i$-butanol in air at $P_C = 15$ bar (squares) and 30 bar (circles).

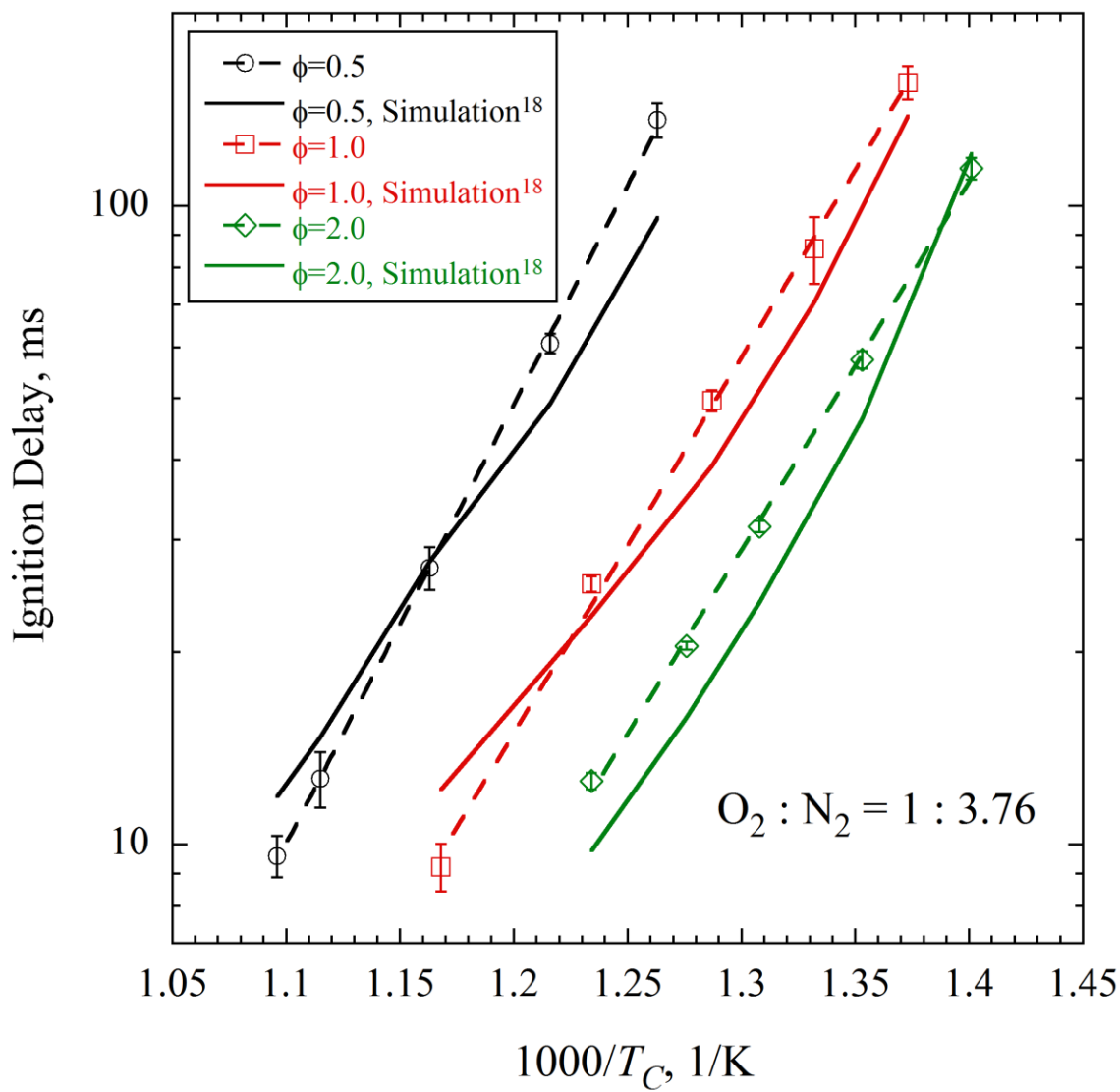

**Figure 12.** Comparison of the simulations using the kinetic mechanism of Sarathy et al.[18] for three equivalence ratio mixtures of t-butanol in air at $P_C = 30$ bar.

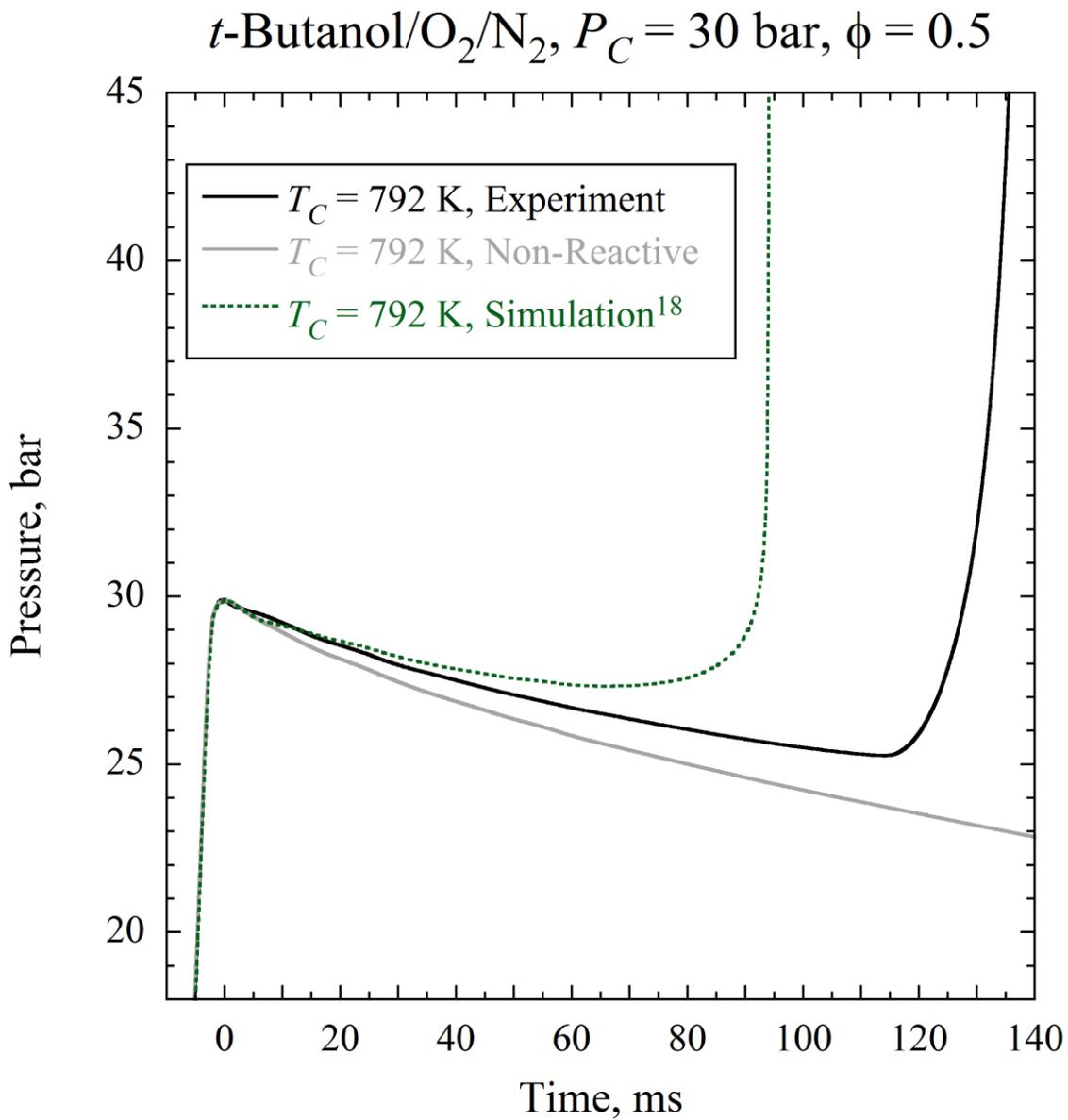

**Figure 13.** $\phi = 0.5$ in air. Pressure traces of selected *t*-butanol experiments compared with the corresponding non-reactive and simulated traces, using the mechanism of Sarathy et al.[18].

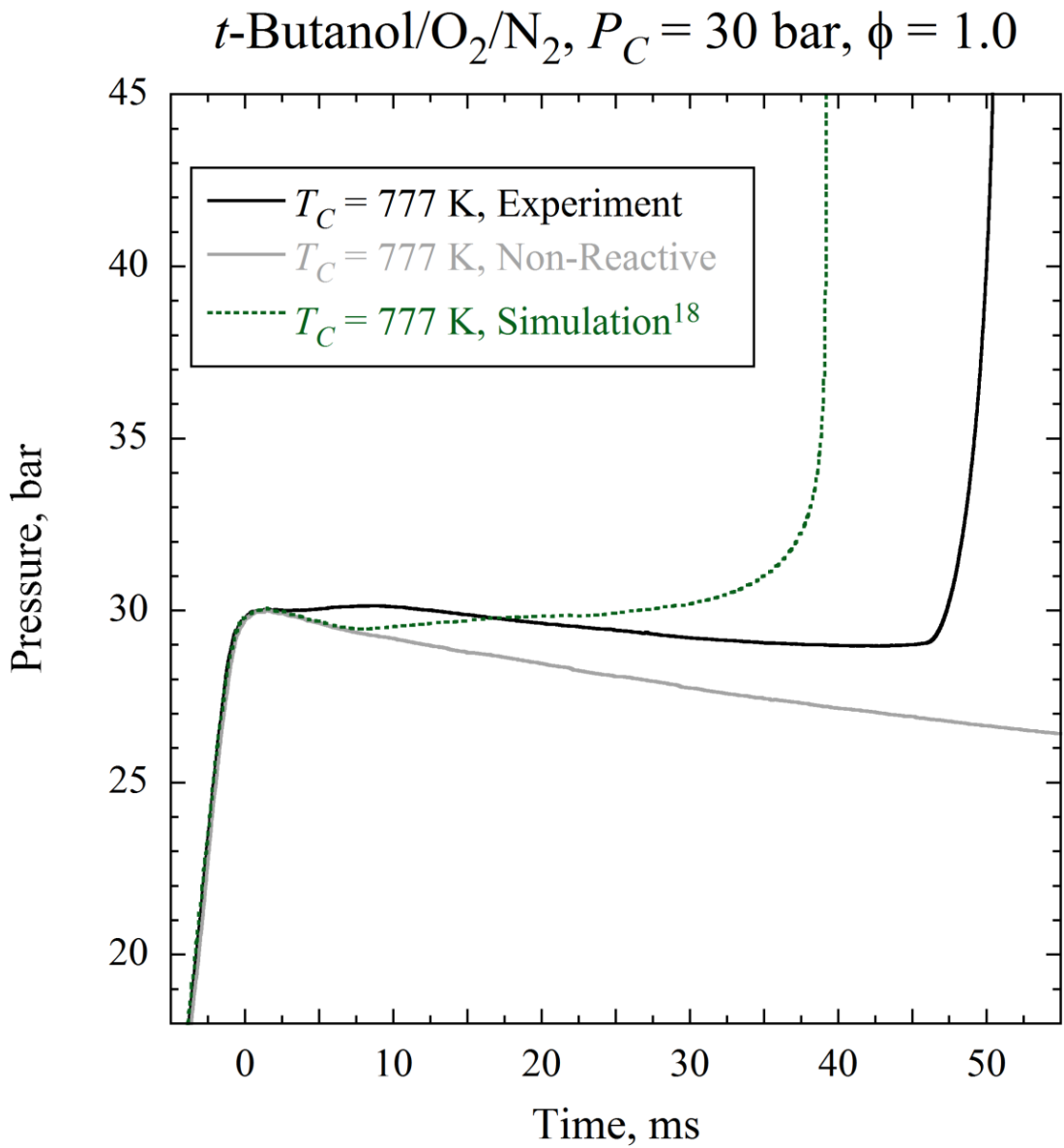

**Figure 14.** $\phi = 1.0$ in air. Pressure traces of selected *t*-butanol experiments compared with the corresponding non-reactive and simulated traces, using the mechanism of Sarathy et al.[18].

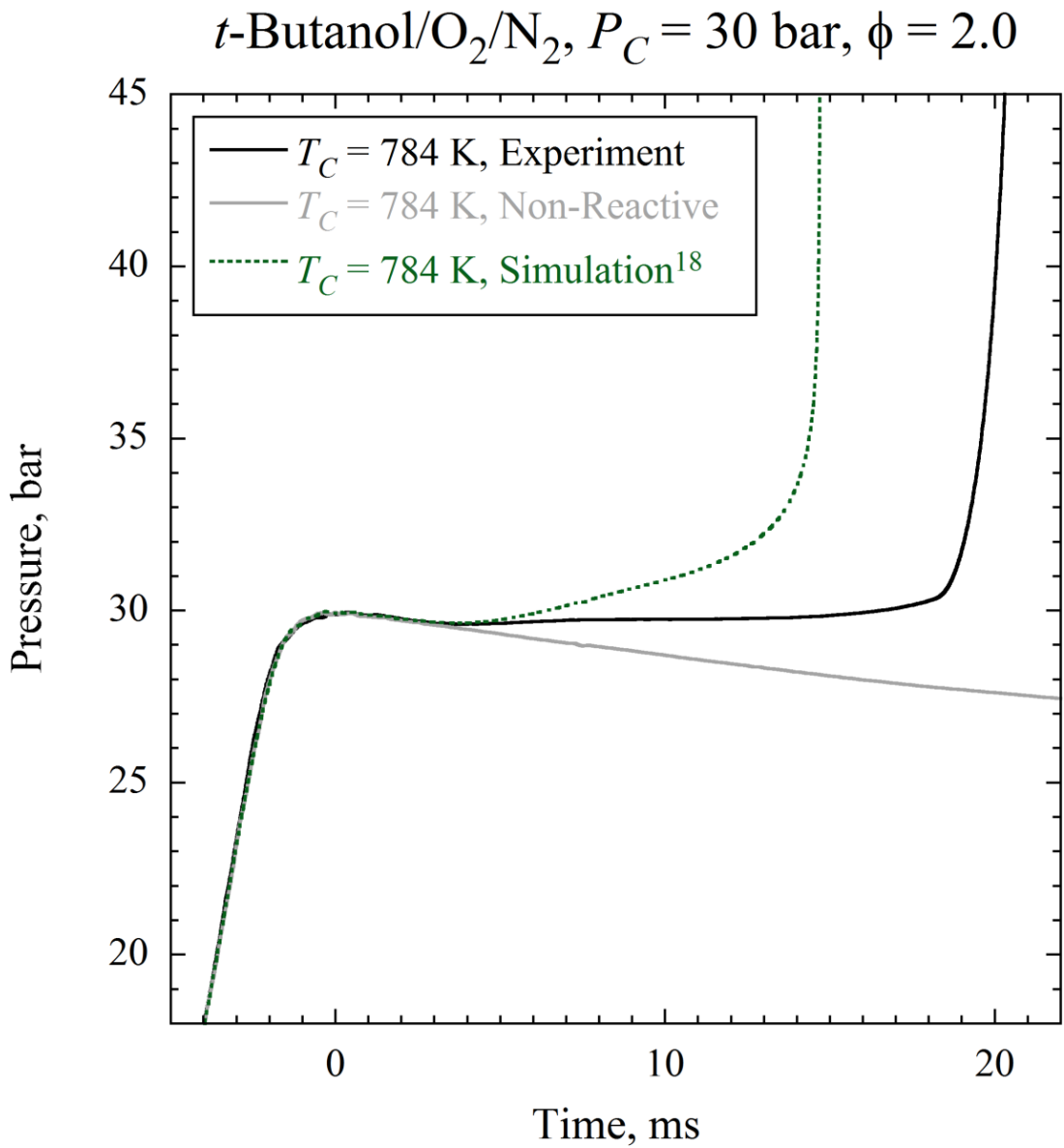

**Figure 15.** $\phi = 2.0$ in air. Pressure traces of selected *t*-butanol experiments compared with the corresponding non-reactive and simulated traces, using the mechanism of Sarathy et al.[18].

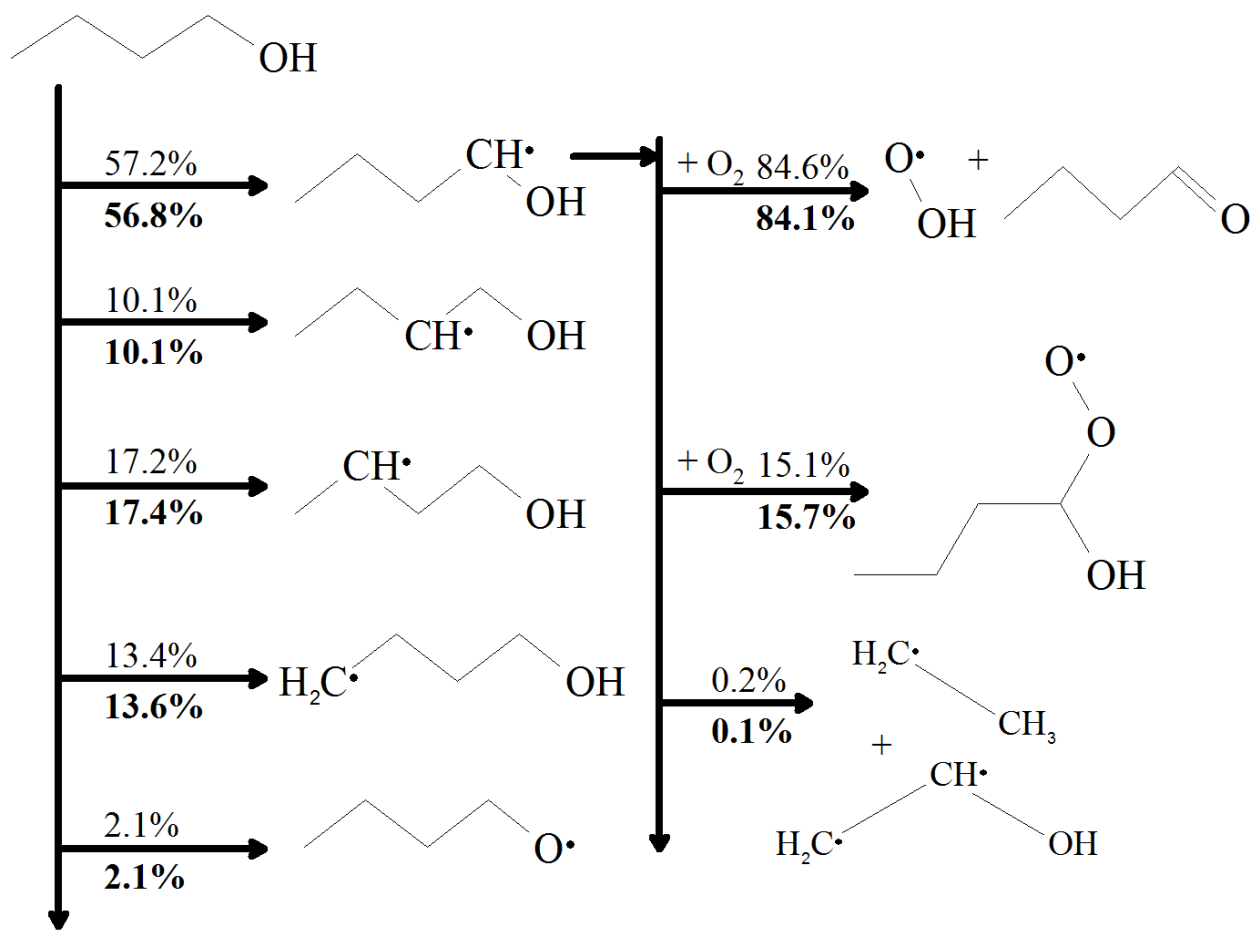

**Figure 16.** Pathway analysis for simulations of *n*-butanol at temperature of 750 K, in stoichiometric air, using the mechanism of Sarathy et al.[18] Percentages in normal text represent an initial condition of 15 bar; bold text is for 30 bar.

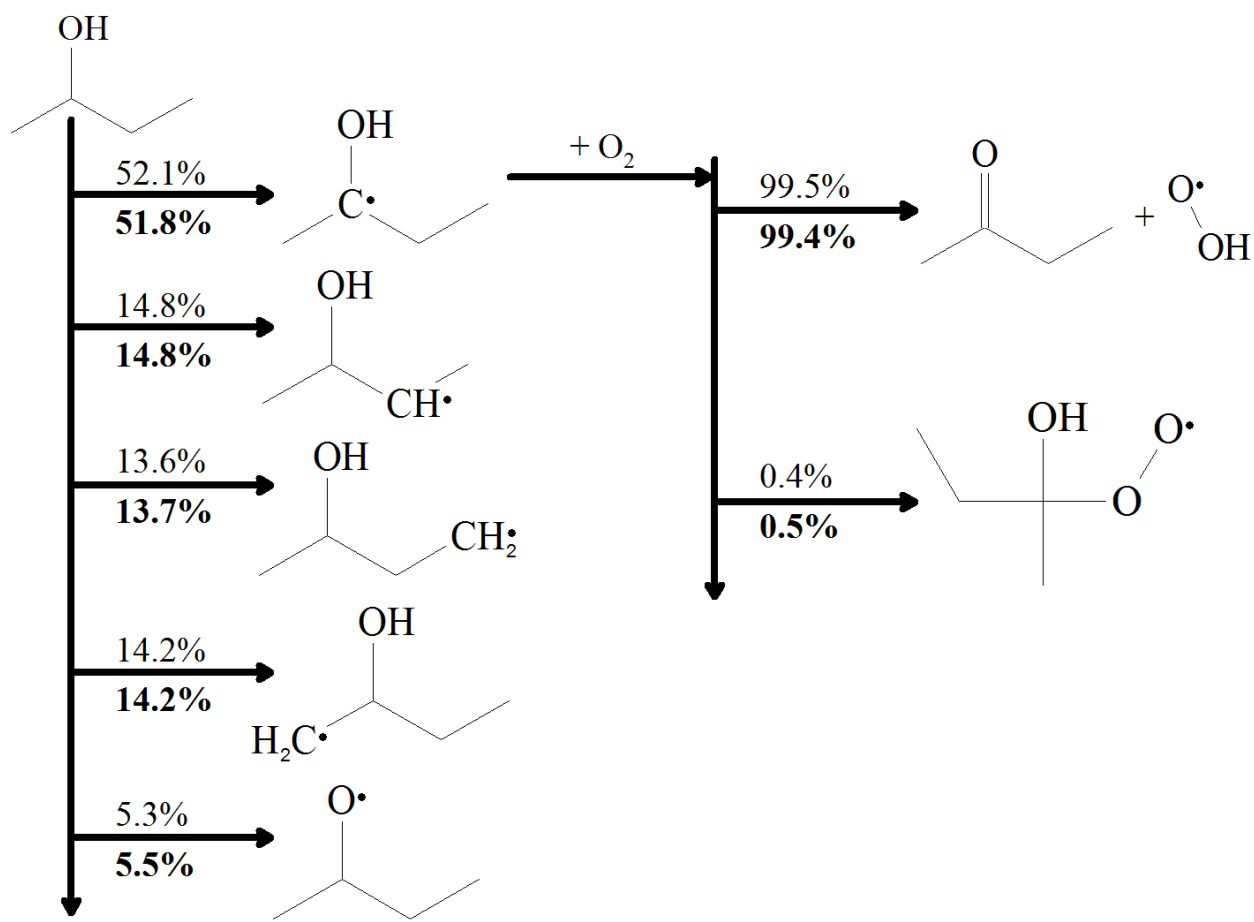

**Figure 17.** Pathway analysis for simulations of *s*-butanol at temperature of 750 K, in stoichiometric air, using the mechanism of Sarathy et al.[18] Percentages in normal text represent an initial condition of 15 bar; bold text is for 30 bar.

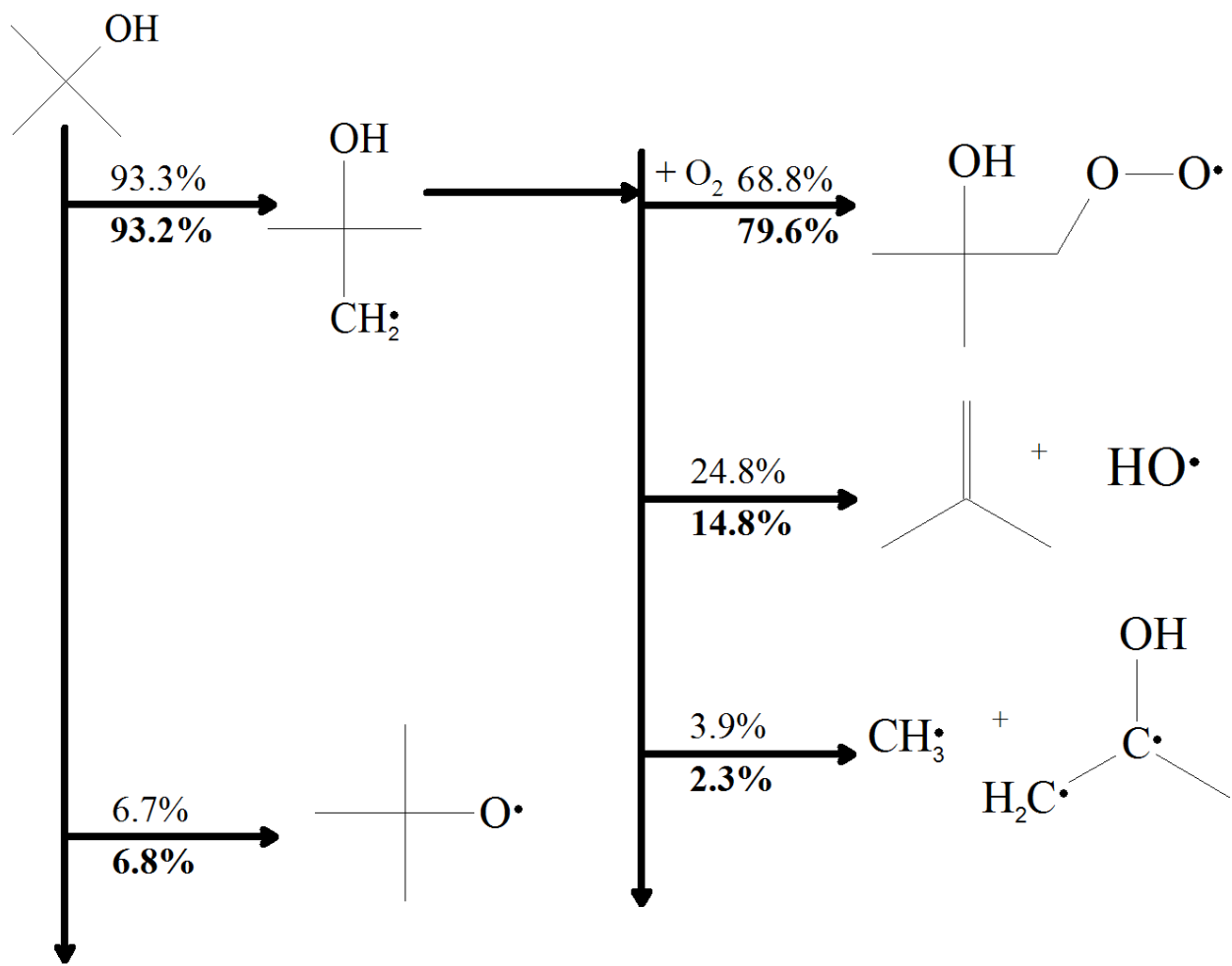

**Figure 18.** Pathway analysis for simulations of *t*-butanol at temperature of 750 K, in stoichiometric air, using the mechanism of Sarathy et al.[18] Percentages in normal text represent an initial condition of 15 bar; bold text is for 30 bar.

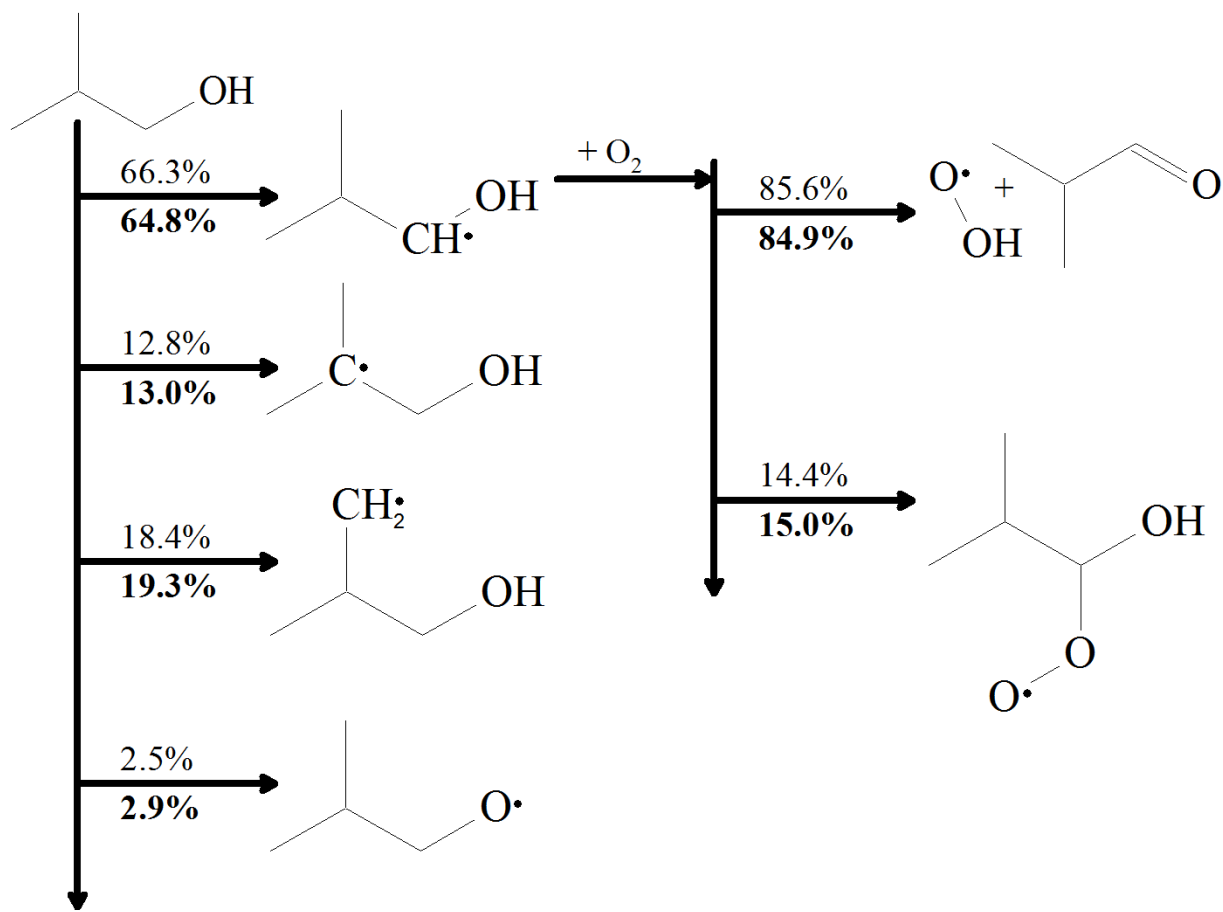

**Figure 19.** Pathway analysis for simulations of *i*-butanol at temperature of 750 K, in stoichiometric air, using the mechanism of Sarathy et al.[18] Percentages in normal text represent an initial condition of 15 bar; bold text is for 30 bar.

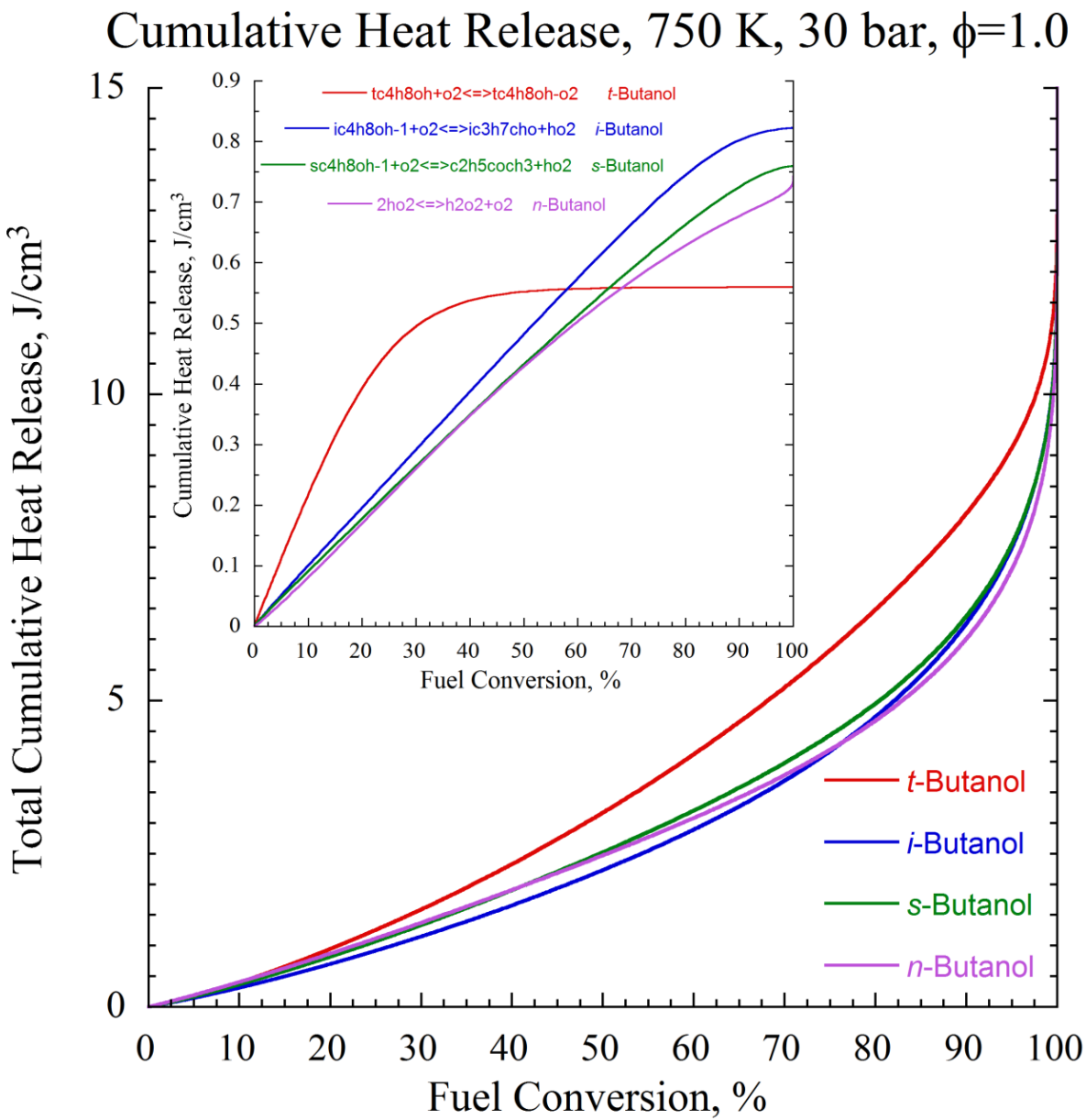

**Figure 20.** Total cumulative heat release and cumulative heat release by important reactions (inset) as a function of fuel consumption from a simulation using the mechanism of Sarathy et al.[18] with initial conditions of 750 K and 30 bar, in stoichiometric air.

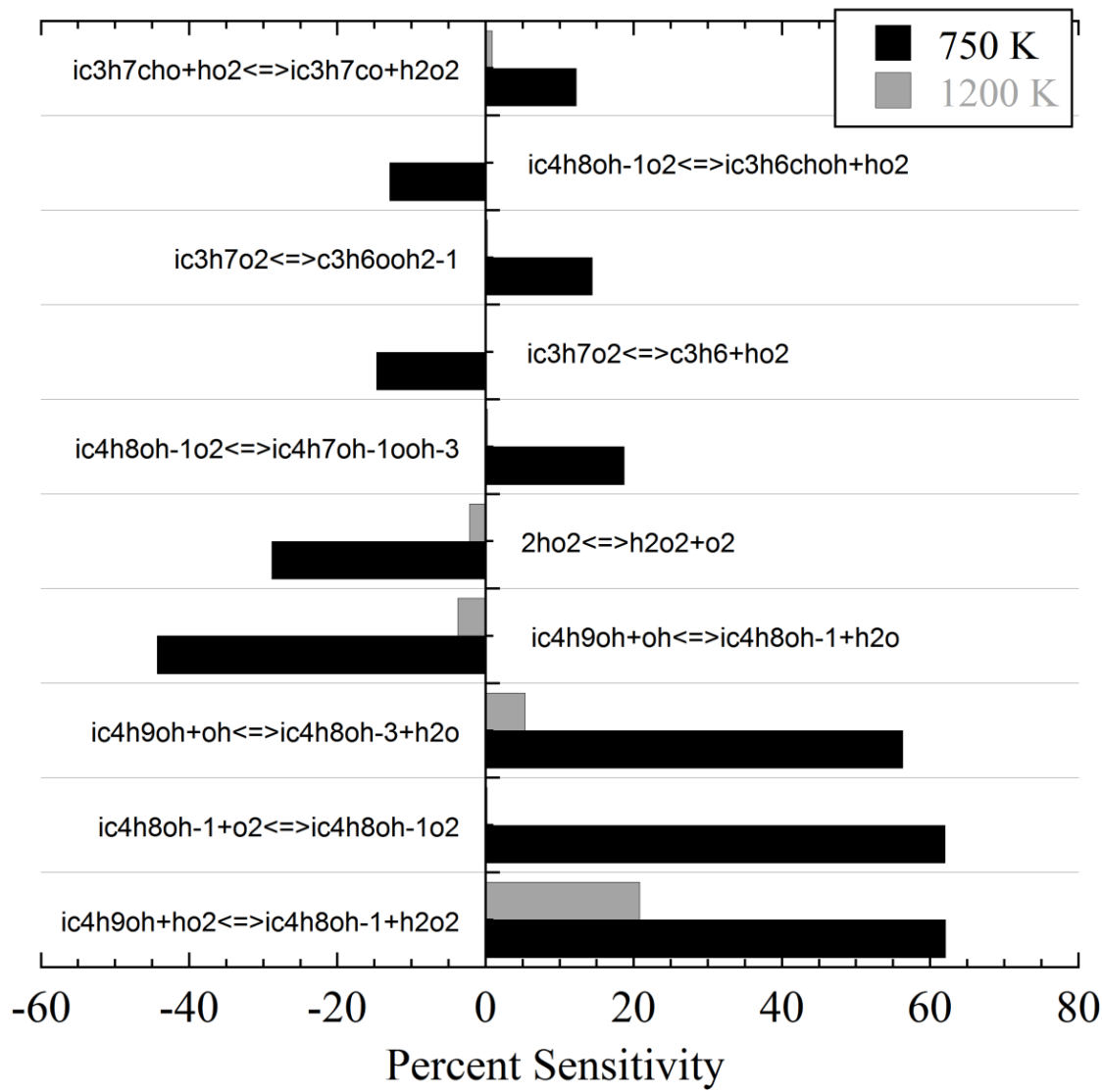

**Figure 21.** Linear brute force sensitivity analysis of the ignition delay with respect to the A-factors of the listed reactions in the mechanism from Sarathy et al.[18] Positive quantities indicate the ignition delay is increased when the A-factor is halved.